\documentclass[
    aps,
    pra,
    twocolumn,
    superscriptaddress,
    amsmath,
    amssymb,
    floatfix
    ]{revtex4-1} 
\usepackage{algorithm}
\usepackage{graphicx} 
\usepackage{hyperref} 
\usepackage{braket}
\usepackage{mathtools}
\usepackage{bm}
\usepackage[normalem]{ulem}
\usepackage{xcolor}
\usepackage{amsthm}
\usepackage{multirow}
\usepackage{algpseudocode, threeparttable}
\usepackage{marvosym}

\begin{document}

\title{Proactively incremental-learning QAOA}

\author{Lingxiao Li}
\affiliation{State Key Laboratory of Networking and Switching Technology, Beijing University of Posts and Telecommunications, Beijing 100876, China.}

\author{Jing Li}
\affiliation{State Key Laboratory of Networking and Switching Technology, Beijing University of Posts and Telecommunications, Beijing 100876, China.}

\author{Yanqi Song}
\affiliation{State Key Laboratory of Networking and Switching Technology, Beijing University of Posts and Telecommunications, Beijing 100876, China.}

\author{Sujuan Qin}
\affiliation{State Key Laboratory of Networking and Switching Technology, Beijing University of Posts and Telecommunications, Beijing 100876, China.}

\author{Qiaoyan Wen}
\affiliation{State Key Laboratory of Networking and Switching Technology, Beijing University of Posts and Telecommunications, Beijing 100876, China.}

\author{Fei Gao\textsuperscript{\Letter}}
\thanks{Corresponding author, gaof@bupt.edu.cn}
\affiliation{State Key Laboratory of Networking and Switching Technology, Beijing University of Posts and Telecommunications, Beijing 100876, China.}

\date{\today}

\begin{abstract}
Solving optimization problems with high performance is the target of existing works of Quantum Approximate Optimization Algorithm (QAOA). With this intention, we propose an advanced QAOA based on incremental learning, where the training trajectory is proactively segmented into incremental phases. Taking the MaxCut problem as our example, we randomly select a small subgraph from the whole graph and train the quantum circuit to get optimized parameters for the MaxCut of the subgraph in the first phase. Then in each subsequent incremental phase, a portion of the remaining nodes and edges are added to the current subgraph, and the circuit is retrained to get new optimized parameters. The above operation is repeated until the MaxCut problem on the whole graph is solved. The key point is that the optimized parameters of the previous phase will be reused in the initial parameters of the current phase. Numerous simulation experiments show our method has superior performance on Approximation Ratio (AR) and training time compared to prevalent works of QAOA. Specifically, the AR is higher than standard QAOA by 13.17\% on weighted random graphs.
\end{abstract}

\maketitle

\section{Introduction}
With the rapid evolution of quantum computing technology \cite{1,2,3}, the advancement of Noisy Intermediate-Scale Quantum (NISQ) \cite{4,5,6} machines has been notable. With the high demand for efficient algorithms for NISQ devices, Variational Quantum Algorithms (VQAs) \cite{7,8,9} are widely studied, which consist of parameterized quantum circuits and classical optimizers. As a quintessential representative of VQAs, the Quantum Approximate Optimization Algorithm (QAOA) offers a novel approach for addressing optimization problems \cite{10,11,12,13,14,15} and exhibits potential quantum advantages compared with classical algorithms  \cite{16,17,18,19,20,21,22}.

Superior performance in solving combinatorial optimization problems is the primary focus for QAOA \cite{23,24,25,26,27,28,29}. In recent years, numerous improved QAOA works have emerged from various perspectives, where initial parameter selection \cite{30,31,32,33,34}, parameter update \cite{35,36,37,38,39,40}, and circuit construction strategy \cite{41,42,43,44} are predominant. Among these improved methods, some works garner widespread attention and show good performance in solving the MaxCut problem, a significant application of QAOA \cite{23,24,25,44,45,46,47,48,49,50,51}. Typically, Medvidović et al. \cite{45} proposed a method for QAOA based on the Restricted Boltzmann Machine (RBM), introducing an efficient approach for simulating quantum circuits without the need for large computing resources and significantly expanding the possibilities to simulate NISQ-era quantum optimization algorithms. Zhou et al. \cite{46} developed the QAOA-in-QAOA(QAOA2) based on the divide-and-conquer heuristic so that a large-scale MaxCut problem can be solved on small quantum machines, providing an efficient strategy for solving large-scale optimization problems and enhancing the applicability of QAOA. Herrman et al. \cite{47} presented the Multi-angle QAOA where a variational ansatz was investigated to reduce circuit depth by increasing the number of classical parameters, increasing the performance of problem-solving with shallower circuits, and showing effective results in the presence of noise. Wurtz et al. \cite{48} the Fixed-angle QAOA, offering a reliable and optimization-free method and yielding an excellent performance on 3-regular graphs. Through different improvement approaches, these works achieved state-of-the-art performances at the time, which promoted the development of QAOA.

Here we improve QAOA from a unique perspective, proposing a novel QAOA based on incremental learning  \cite{52,53,54,55}, called Proactively-Incremental-Learning QAOA (PIL QAOA). The incremental-learning-based methods have outstanding performance in classical deep-learning \cite{56,57,58} for updating parameters. So, this training paradigm is introduced into PIL QAOA. Specifically, the focus of PIL QAOA is that the training trajectory is proactively segmented into incremental phases. PIL QAOA starts with an easily solvable MaxCut problem on a subgraph, then gradually extends this subgraph and retrains the circuit using previous parameters until the Maxcut problem on the target graph is solved. As a result, simulation experiments show that PIL QAOA has outstanding solution ability and less training time than prevalent works of QAOA. Our performance on approximation ratio (AR) is higher by 10.36\% and training time is reduced by 71.43\% than that of Standard QAOA. At the same time, PIL QAOA is also better than other algorithms in terms of anti-forgetting and stability.

The main contributions of this paper are shown below:

\begin{figure*}[!htbp]
	\includegraphics[width=0.99\textwidth]{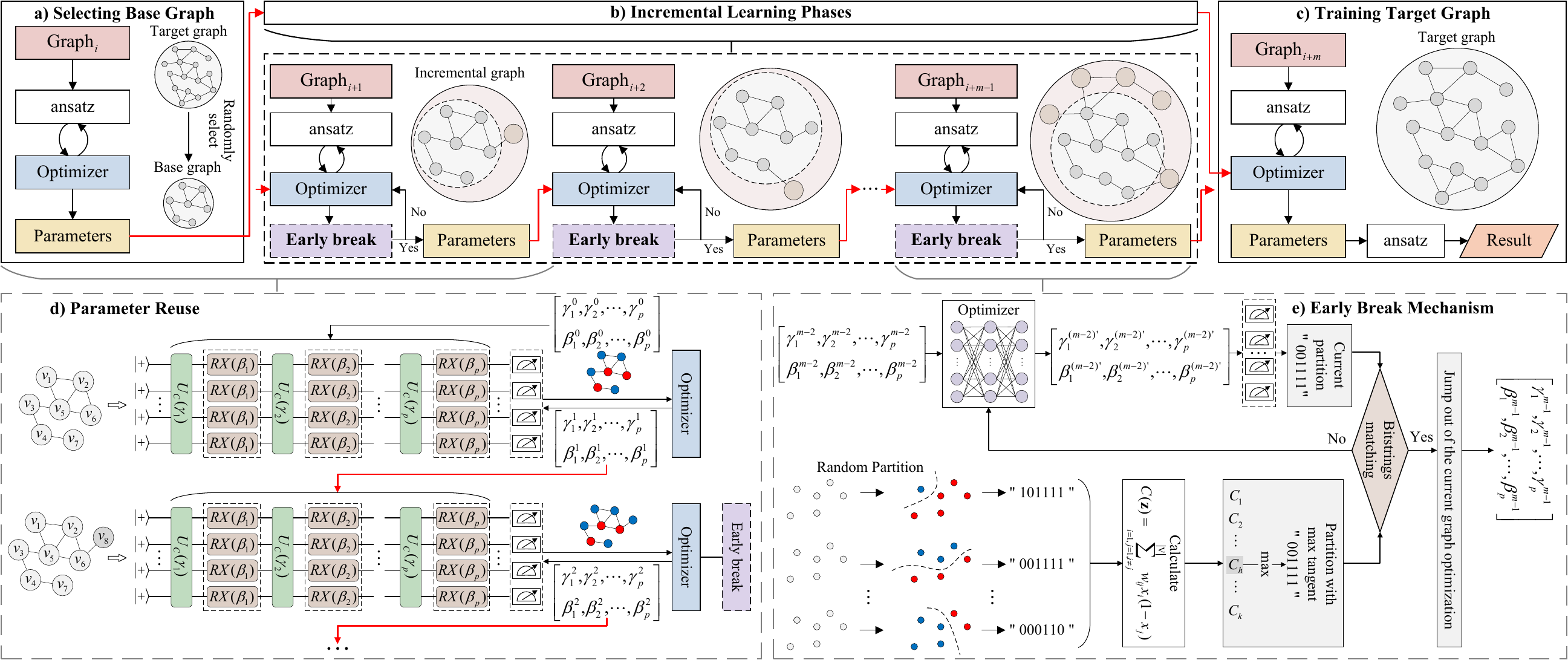}
	\caption{\small{\textbf{Overall framework of the PIL QAOA method for solving the MaxCut problem.} The top parts \textbf{a)}, \textbf{b)}, and \textbf{c)} show the overview, while the bottom parts \textbf{d)} and \textbf{e)} show the design details of "parameter reuse" and "early break mechanism" in detail. \textbf{a)} Training the base graph which is randomly selected from the target graph, and the optimized parameters are obtained. \textbf{b)} The base graph is incremental by adding a portion of the remaining nodes and edges, and the initial parameters of the current phase are set as the optimized parameters of the previous phase. This part includes the "parameter reuse" and "early break mechanism". \textbf{c)} Training the target graph to get the ultimate optimized parameters. \textbf{d)} The red arrow signifies the process of parameter reuse. \textbf{e)} When the currently optimized cut value is greater or equal to the maximum value of multiple random divisions, there is no need to wait for convergence. Then the next training phase is started, and the current optimized parameters are reused in the initial parameters of the next phase.}}
	\label{fig1}
\end{figure*}

\begin{enumerate}
	\item Proactively incremental learning: PIL QAOA is a method to solve optimization problems based on the idea of incremental learning. Taking the MaxCut problem as our example, a small subgraph (called base graph) is randomly selected from the whole graph (called target graph) in the first phase, and training the quantum circuit to get optimized parameters that imply its maximum cut. The MaxCut problem on the base graph is simpler to solve than that of the target graph. Then a portion of the remaining nodes and edges are added to the base graph to generate an incremental graph, and the circuit is retrained to get new optimized parameters. Next, in each subsequent phase, a portion of the remaining is further added to the incremental graph, and the circuit is retrained. The initial parameters of the current phase are reused from the optimized parameters in the previous phase. The adding, reusing, and retraining process continues until the MaxCut problem on the target graph is solved.
	
	\item Early break mechanism: In order to further shorten the training time of the incremental phase, we design the "early break" mechanism. When the optimized cut value is not less than the maximum cut value in k random partitions, the optimization of the next phase can be processed without waiting for convergence.
	
	\item Extensive analysis of simulation experiments: To fully verify the performance of methods, we construct the MaxCut-Sandbox dataset. The dataset contains weighted graphs with three categories (random, regular, and complete), and unweighted graphs with three categories (random, regular, and complete). The number of nodes for each graph ranges from 5 to 10. The comparison experiment with existing QAOAs is composed of the performance on AR, training time, degree of anti-forgetting, and solving stability. 
\end{enumerate}

\section{Result}
\subsection{PIL QAOA}

Most works of QAOA focus on the MaxCut problem, so PIL QAOA also elaborates on this problem. Existing works of QAOA usually directly train all nodes and edges of the graph when solving. The solution space tends to grow exponentially as the number of nodes increases. Solving in such a large solution space often results in poor performance. PIL QAOA adopts a progressive approach from a unique perspective, commencing with a smaller subgraph (called base graph) and expanding its scale gradually. Reusing previously optimized parameters by training to offer potent direction for subsequent incremental phases. The method implements an incremental way of training, which increases the performance. The overall framework process of PIL QAOA is illustrated in FIG. \ref{fig1}.

\begin{figure*}[!htbp]
	\includegraphics[width=0.90\textwidth]{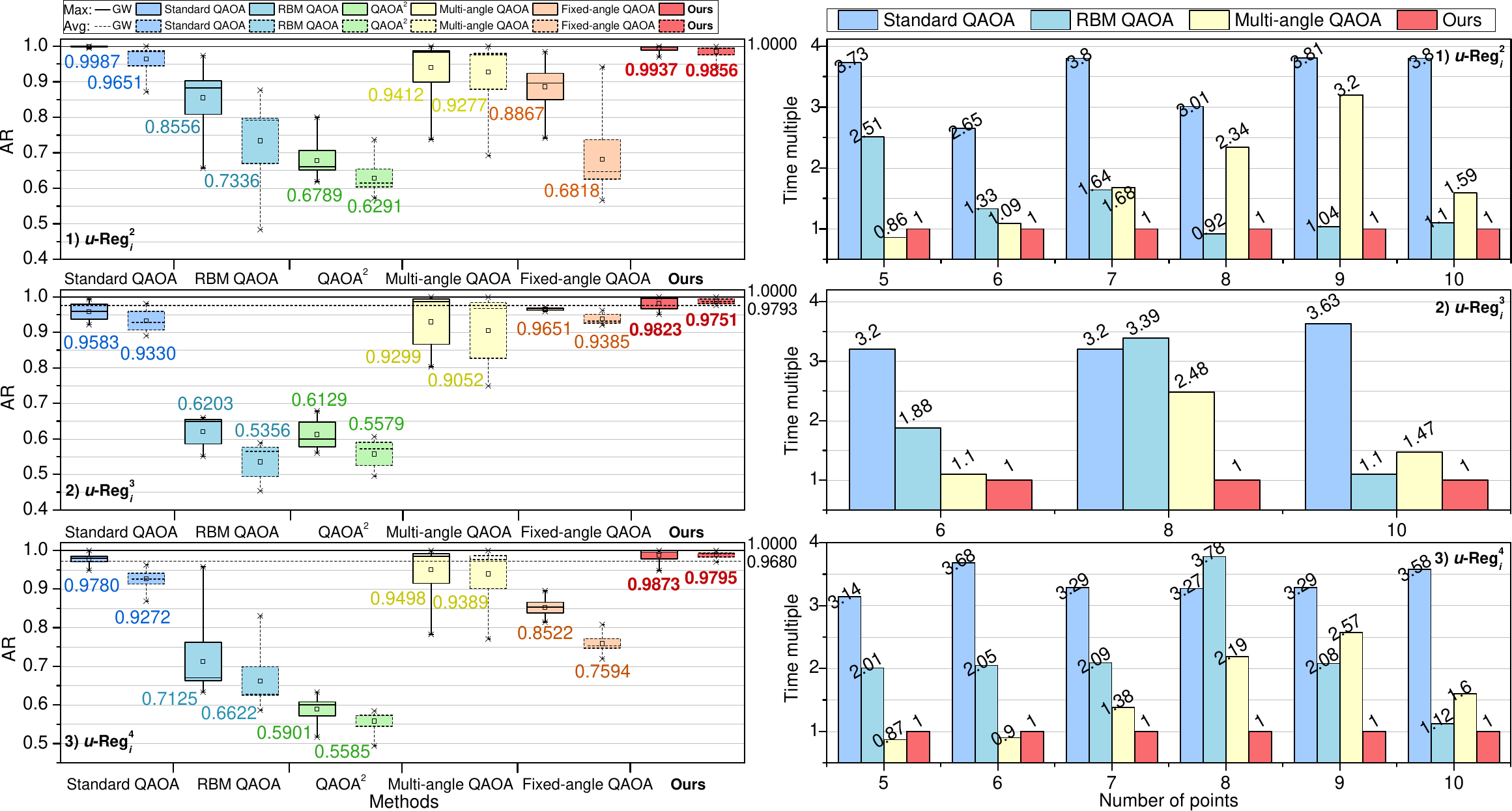}
	\caption{\small{\textbf{Comparison of AR and training time between the baseline methods and the proposed method in unweighted regular graphs.} The left subfigure is the AR comparison, and the right subfigure is the training time comparison. In the left subfigure, each box is the overall AR values of the method in multiple graphs. The solid box and the dotted box represent the max and average distribution of AR in these graphs of the method. For example, in subfigure 1), the box corresponding to 0.9987 is the overall max AR of Standard QAOA on the $u\mathrm{-Reg}^2_i$ graphs ($i=5,6,7,8,9,10$), and the box corresponding to 0.9651 is the overall average AR of Standard QAOA on the $u\mathrm{-Reg}^2_i$ graphs ($i=5,6,7,8,9,10$). The horizontal axis represents different methods, and the vertical axis represents the AR value. The solid and dotted lines represent the max and average AR of GW. The specific number of each box is the average value of these ARs in multiple graphs. In the right subfigure, the specific number of each bar represents the training time multiple of the methods compared to PIL QAOA.}}
	\label{fig2}
\end{figure*}

\subsection{Comparison experiments on MaxCut}
As mentioned in the Introduction, several quantum methods including Standard QAOA \cite{10}, RBM QAOA \cite{45}, $\mathrm{QAOA}^2$ \cite{46}, Multi-angle QAOA \cite{47}, and Fixed-angle QAOA \cite{48}, have shown promising performance and offered valuable insights into the training of QAOA. Thus, this paper uses them as the baseline methods to conduct a comparison analysis with PIL QAOA proposed. Besides, GW \cite{59}, as a representative classical algorithm, is also considered. We "reproduce" these methods for selected problem instances and compare the AR and training time on six types of graphs (unweighted: regular, random, and complete graphs; weighted: regular, random, and complete graphs) with a 3-layers QAOA circuit.  We give this difference in Time multiple $T_x/T_{ours}$, where $T_x$ represents the training time of various baseline methods, and $T_{ours}$ represents the complete time from selecting the base graph to completing the target graph optimization. Next, the stability of AR and the degree of anti-forgetting are compared. Furthermore, we also explore the selection of $p$-layers within the PIL QAOA method.

The comparison results in unweighted regular graphs and weighted random graphs are shown in the main paper, because regular graphs are commonly used in experiments of unweighted graphs \cite{23,24,42}, and weighted random graphs are known for their complexity in solving. The comparison of other types of graphs (unweighted: random and complete graphs; weighted: regular and complete graphs) is shown in the Supplementary Information. 

To ensure the fairness and reliability of our experiments, each experiment was repeated ten times for every instance, and the results were subsequently averaged to provide an evaluation.

\subsubsection{Unweighted graphs}
On unweighted regular graphs, FIG. \ref{fig2} illustrates the AR and training time comparison between the baseline methods and PIL QAOA method. The experimental results on other types of unweighted graphs are shown in Supplementary information.

\begin{table*}[!ht]
	\centering
	\fontsize{6pt}{7pt}\selectfont
	\setlength{\tabcolsep}{0.5mm}
	\caption{\small{\textbf{Comparison of AR and training time between the baseline methods and the proposed method in unweighted regular graphs.} Each experiment was repeated ten times for every graph, and the results were averaged to provide. \textcolor{gray}{Gray} represents the performance of the classical GW method. \textbf{Bold} represents the best AR value of these QAOA methods.}}\label{tab1}
	\begin{tabular}{|c|cccccc|c|ccc|c|cccccc|c|}
		\hline
		\textbf{\multirow{2}*{AR (max)}} & \multicolumn{7}{c|}{$d=2$} & \multicolumn{4}{c|}{$d=3$} & \multicolumn{7}{c|}{$d=4$} \\ \cline{2-19}
		\textbf{} & $i=5$ & $i=6$ & $i=7$ & $i=8$ & $i=9$ & $i=10$ & Avg ($i$) & $i=6$ & $i=8$ & $i=10$ & Avg ($i$) & $i=5$ & $i=6$ & $i=7$ & $i=8$ & $i=9$ & $i=10$ & Avg ($i$) \\ \hline
		\textcolor{gray}{GW (Classical)} & \textcolor{gray}{1.0000} & \textcolor{gray}{1.0000} & \textcolor{gray}{1.0000} & \textcolor{gray}{1.0000} & \textcolor{gray}{1.0000} & \textcolor{gray}{1.0000} & \textcolor{gray}{1.0000} & \textcolor{gray}{1.0000} & \textcolor{gray}{1.0000} & \textcolor{gray}{1.0000} & \textcolor{gray}{1.0000} & \textcolor{gray}{1.0000} & \textcolor{gray}{1.0000} & \textcolor{gray}{1.0000} & \textcolor{gray}{1.0000} & \textcolor{gray}{1.0000} & \textcolor{gray}{1.0000} & \textcolor{gray}{1.0000} \\ \hline
		Standard QAOA & 1.0000 & 1.0000 & 1.0000 & 1.0000 & 0.9975 & 0.9947 & \textbf{0.9987} & 0.9943 & 0.9590 & 0.9215 & 0.9583 & 1.0000 & 0.9900 & 0.9860 & 0.9733 & 0.9486 & 0.9700 & 0.9780 \\ 
		RBM QAOA & 0.9201 & 0.9734 & 0.9342 & 0.8451 & 0.8039 & 0.6570 & 0.8556 & 0.5512 & 0.6601 & 0.6496 & 0.6203 & 0.9575 & 0.6656 & 0.6750 & 0.7047 & 0.6389 & 0.6335 & 0.7125 \\ 
		$\mathrm{QAOA}^2$ & 0.8000 & 0.7000 & 0.6889 & 0.6328 & 0.6328 & 0.6187 & 0.6789 & 0.6786 & 0.5600 & 0.6000 & 0.6129 & 0.6282 & 0.6333 & 0.5625 & 0.5167 & 0.5982 & 0.6016 & 0.5901 \\ 
		Multi-angle QAOA & 1.0000 & 1.0000 & 0.9977 & 0.9757 & 0.9358 & 0.7377 & 0.9412 & 0.9999 & 0.9867 & 0.8032 & 0.9299 & 1.0000 & 1.0000 & 0.9980 & 0.9843 & 0.9337 & 0.7830 & 0.9498 \\ 
		Fixed-angle QAOA & 0.9277 & 0.9844 & 0.8658 & 0.9650 & 0.8349 & 0.7421 & 0.8867 & 0.9710 & 0.9655 & 0.9589 & 0.9651 & 0.8955 & 0.8725 & 0.8332 & 0.8206 & 0.8771 & 0.8141 & 0.8522 \\ 
		\textbf{Ours} & 1.0000 & 0.9993 & 0.9957 & 1.0000 & 0.9970 & 0.9910 & 0.9972 & 0.9998 & 0.9960 & 0.9900 & \textbf{0.9953} & 1.0000 & 1.0000 & 0.9977 & 0.9961 & 0.9981 & 0.9851 & \textbf{0.9962} \\ \hline
		
		\textbf{\multirow{2}*{AR (avg)}} & \multicolumn{7}{c|}{$d=2$} & \multicolumn{4}{c|}{$d=3$} & \multicolumn{7}{c|}{$d=4$} \\ \cline{2-19}
		\textbf{} & $i=5$ & $i=6$ & $i=7$ & $i=8$ & $i=9$ & $i=10$ & Avg ($i$) & $i=6$ & $i=8$ & $i=10$ & Avg ($i$) & $i=5$ & $i=6$ & $i=7$ & $i=8$ & $i=9$ & $i=10$ & Avg ($i$) \\ \hline
		GW (Classical) & \textcolor{gray}{1.0000} & \textcolor{gray}{1.0000} & \textcolor{gray}{1.0000} & \textcolor{gray}{1.0000} & \textcolor{gray}{1.0000} & \textcolor{gray}{1.0000} & \textcolor{gray}{1.0000} & \textcolor{gray}{1.0000} & \textcolor{gray}{0.9745} & \textcolor{gray}{0.9634} & \textcolor{gray}{0.9793} & \textcolor{gray}{0.9733} & \textcolor{gray}{0.9655} & \textcolor{gray}{1.0000} & \textcolor{gray}{0.9634} & \textcolor{gray}{0.9335} & \textcolor{gray}{0.9725} & \textcolor{gray}{0.9680} \\ \hline
		Standard QAOA & 0.9980 & 1.0000 & 0.9960 & 0.9780 & 0.9465 & 0.8720 & 0.9651 & 0.9806 & 0.9280 & 0.8905 & 0.9330 & 0.9293 & 0.9215 & 0.9624 & 0.9607 & 0.9206 & 0.8688 & 0.9272 \\ 
		RBM QAOA & 0.8766 & 0.8371 & 0.8460 & 0.7462 & 0.4836 & 0.6118 & 0.7336 & 0.4540 & 0.5883 & 0.5644 & 0.5356 & 0.8304 & 0.5866 & 0.6074 & 0.6947 & 0.6316 & 0.6224 & 0.6622 \\ 
		$\mathrm{QAOA}^2$ & 0.7371 & 0.6587 & 0.6152 & 0.5725 & 0.6138 & 0.5770 & 0.6291 & 0.6063 & 0.4954 & 0.5719 & 0.5579 & 0.5749 & 0.5843 & 0.5438 & 0.4938 & 0.5812 & 0.5727 & 0.5585 \\ 
		Multi-angle QAOA & 1.0000 & 1.0000 & 0.9966 & 0.9642 & 0.9131 & 0.6921 & 0.9277 & 0.9994 & 0.9674 & 0.7489 & 0.9052 & 1.0000 & 0.9999 & 0.9978 & 0.9748 & 0.8897 & 0.7711 & 0.9389 \\ 
		Fixed-angle QAOA & 0.6997 & 0.5899 & 0.6605 & 0.941 & 0.6332 & 0.5662 & 0.6818 & 0.9622 & 0.9212 & 0.9321 & 0.9385 & 0.7733 & 0.8078 & 0.7537 & 0.7525 & 0.7191 & 0.7501 & 0.7594 \\ 
		\textbf{Ours} & 0.9993 & 0.9985 & 0.9910 & 0.998 & 0.9860 & 0.9810 & \textbf{0.9923} & 0.9991 & 0.9872 & 0.9758 & \textbf{0.9874} & 1.0000 & 0.9977 & 0.9986 & 0.9857 & 0.9698 & 0.9791 & \textbf{0.9885} \\ \hline
		
		\textbf{\multirow{2}*{Time (second)}} & \multicolumn{7}{c|}{$d=2$} & \multicolumn{4}{c|}{$d=3$} & \multicolumn{7}{c|}{$d=4$} \\ \cline{2-19}
		\textbf{} & $i=5$ & $i=6$ & $i=7$ & $i=8$ & $i=9$ & $i=10$ & Avg ($i$) & $i=6$ & $i=8$ & $i=10$ & Avg ($i$) & $i=5$ & $i=6$ & $i=7$ & $i=8$ & $i=9$ & $i=10$ & Avg ($i$) \\ \hline
		Standard QAOA & 45.66 & 44.46 & 97.43 & 115.70 & 319.83 & 935.44 & 259.75 & 62.22 & 128.04 & 958.51 & 382.92 & 54.80 & 86.76 & 101.61 & 150.08 & 305.09 & 962.79 & 276.86 \\ 
		RBM QAOA & 30.66 & 22.39 & 42.14 & 35.35 & 87.59 & 270.27 & 81.40 & 36.63 & 135.76 & 290.85 & 154.41 & 35.20 & 48.25 & 64.65 & 173.89 & 193.07 & 300.20 & 135.88 \\ 
		Multi-angle QAOA & 10.46 & 18.24 & 43.14 & 90.00 & 268.84 & 391.49 & 137.03 & 21.35 & 99.40 & 387.25 & 169.33 & 15.18 & 21.26 & 42.71 & 100.67 & 238.30 & 429.64 & 141.29 \\ 
		\textbf{Ours} & 12.23 & 16.78 & 25.66 & 38.38 & 83.98 & 246.09 & \textbf{70.52} & 19.46 & 40.04 & 264.01 & \textbf{107.84} & 17.47 & 23.56 & 30.93 & 45.96 & 92.73 & 268.98 & \textbf{79.94} \\ \hline
	\end{tabular}
\end{table*}

Detailed experimental data can be found in TABLE \ref{tab1}. In terms of AR, we select complete baseline methods. In terms of training time, it should be noted that GW is a classical method, the parameters of Fixed-angle QAOA are determined and do not require further training, and $\mathrm{QAOA}^2$ is designed to solve the large-scale MaxCut problem, and its performance on the small-scale graphs is unsatisfactory. Therefore, in this training time comparison, we select the following baseline methods: Standard QAOA, RBM QAOA, and Multi-angle QAOA. It should be noted that each value is the average of 10 test results, and there is rounding.

According to the results in FIG. \ref{fig2}, TABLE \ref{tab1}, and Supplementary TABLE \ref{tab4}, Supplementary TABLE \ref{tab5}, we find that in terms of the AR of unweighted graphs:

\begin{enumerate}
	\item For example, in the 3-regular graphs ($u\mathrm{-Reg}^3_i$), the ARs of PIL QAOA are higher by 37.5\% (max) and 45.18\% (avg) than those of RBM QAOA, and also outperform Fixed-angle QAOA by 3.02\% (max) and 4.89\% (avg). For all unweighted graphs, the ARs of PIL QAOA are higher by about 1.84\%~43.24\%(max) and 3.12\%-43.24\% (avg) than baseline methods.
	\item For all unweighted graphs with less than 10 nodes, the ARs of PIL QAOA decrease relatively slightly as the number of nodes increases. Although Standard QAOA and Multi-angle QAOA perform well in graphs with fewer nodes, their efficiency drops significantly as the number of nodes increases. In contrast, PIL QAOA maintains relatively stable high performance even as the number of nodes increases. At the same time, PIL QAOA displays equivalent solution performance to that of the GW method and even exceeds it in certain cases.
\end{enumerate}

In terms of training time of unweighted graphs:

\begin{enumerate}
	\item PIL QAOA consistently has a shorter training time than the baseline methods in different graphs. For unweighted graphs, the training time of PIL QAOA is reduced by about  32.59\%~64.82\% compared to baseline methods. It can be seen that PIL QAOA greatly improves computational efficiency for various graphs.
\end{enumerate}

\subsubsection{Weighted graphs}
\begin{figure*}[hb]
	\includegraphics[width=0.90\textwidth]{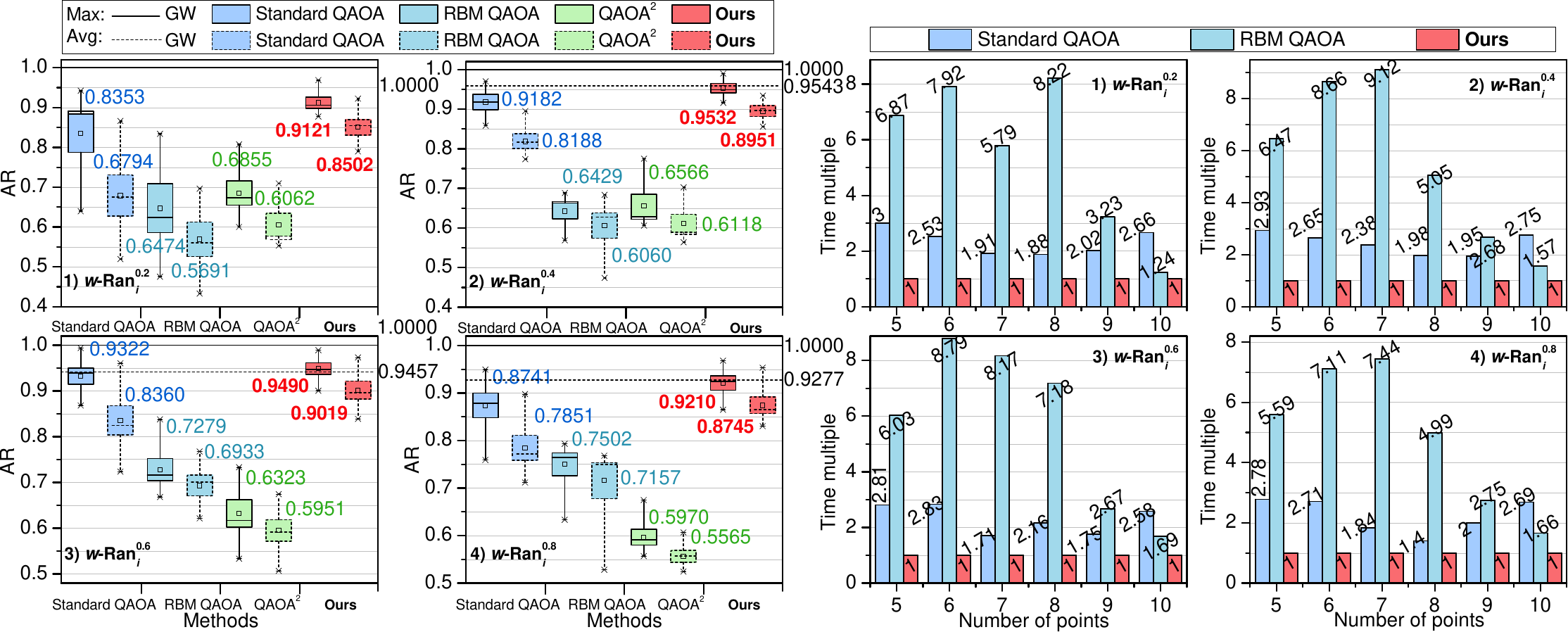}
	\caption{\small{\textbf{Comparison of AR and training time between the baseline methods and the proposed method in weighted random graphs.} The left subfigure is the AR comparison, and the right subfigure is the Time comparison. Each box is the overall AR values of the method in multiple graphs. The solid box and the dotted box represent the max and average AR distribution of AR in these graphs of the method. The horizontal axis represents different methods, and the vertical axis represents the AR value. The solid and dotted lines represent the max and average AR of GW. The specific number of each box is the average value of these ARs in multiple graphs.}}
	\label{fig3}
\end{figure*}

On weighted random graphs, FIG. \ref{fig3} illustrates the AR and training time comparison between the baselines and PIL QAOA method. The experimental results on other types of weighted graphs are shown in Supplementary information. 

Detailed experimental data can be found in TABLE \ref{tab2} It should be noted that Fixed-angle QAOA and Multi-angle QAOA are not suitable for solving weighted graphs, therefore other methods (GW, Standard QAOA, RBM QAOA, and $\mathrm{QAOA}^2$) are employed in this experiment. As for training time, similar to the experiments on unweighted graphs, here we don’t consider $\mathrm{QAOA}^2$ as well, and the selected baseline methods are Standard QAOA and RBM QAOA.

\begin{table*}[bp]
	\centering
	\caption{\small{\textbf{Comparison of AR and training time between the baseline methods and the proposed method in weighted random graphs.} Each experiment was repeated ten times for every graph, and the results were averaged to provide. \textcolor{gray}{Gray} represents the performance of the classical GW method. \textbf{Bold} represents the best AR value of these QAOA methods.}}\label{tab2}
	\tiny
	\setlength{\tabcolsep}{0.7mm}
	\begin{tabular}{|c|cccc|c|cccc|c|cccc|c|}
		\hline
		\textbf{\multirow{2}*{AR (max)}} & \multicolumn{5}{c|}{$i=5$} & \multicolumn{5}{c|}{$i=6$} & \multicolumn{5}{c|}{$i=7$} \\ \cline{2-16}
		~ & $ep=0.2$ & $ep=0.4$ & $ep=0.6$ & $ep=0.8$ & Avg ($ep$)& $ep=0.2$ & $ep=0.4$ & $ep=0.6$ & $ep=0.8$ & Avg ($ep$) & $ep=0.2$ & $ep=0.4$ & $ep=0.6$ & $ep=0.8$ & Avg ($ep$) \\ \hline
		\textcolor{gray}{GW (Classical)} & \textcolor{gray}{1.0000} & \textcolor{gray}{1.0000} & \textcolor{gray}{1.0000} & \textcolor{gray}{1.0000} & \textcolor{gray}{1.0000} & \textcolor{gray}{1.0000} & \textcolor{gray}{1.0000} & \textcolor{gray}{1.0000} & \textcolor{gray}{1.0000} & \textcolor{gray}{1.0000} & \textcolor{gray}{1.0000} & \textcolor{gray}{1.0000} & \textcolor{gray}{1.0000} & \textcolor{gray}{1.0000} & \textcolor{gray}{1.0000} \\ \hline
		Standard QAOA & 0.9035 & 0.9700 & 0.9938 & 0.9135 & 0.9452 & 0.942 & 0.9667 & 0.9551 & 0.9496 & 0.9534 & 0.7452 & 0.8796 & 0.9338 & 0.7595 & 0.8295 \\
		RBM QAOA & 0.8337 & 0.5964 & 0.7041 & 0.6333 & 0.6919 & 0.7956 & 0.6711 & 0.8381 & 0.7932 & 0.7745 & 0.7033 & 0.5686 & 0.6680 & 0.7598 & 0.6749 \\
		$\mathrm{QAOA}^2$ & 0.8077 & 0.7750 & 0.7031 & 0.5938 & 0.7199 & 0.7333 & 0.7083 & 0.7333 & 0.6750 & 0.7125 & 0.6735 & 0.6190 & 0.5903 & 0.5568 & 0.6099 \\
		\textbf{Ours} & 0.9330 & 0.9899 & 0.9893 & 0.9447 & \textbf{0.9642} & 0.9687 & 0.9834 & 0.9710 & 0.9680 & \textbf{0.9728} & 0.8991 & 0.9472 & 0.9534 & 0.8651 & \textbf{0.9162} \\ \hline
		
		\textbf{\multirow{2}*{AR (max)}} & \multicolumn{5}{c|}{$i=8$} & \multicolumn{5}{c|}{$i=9$} & \multicolumn{5}{c|}{$i=10$} \\ \cline{2-16}
		~ & $ep=0.2$ & $ep=0.4$ & $ep=0.6$ & $ep=0.8$ & Avg ($ep$) & $ep=0.2$ & $ep=0.4$ & $ep=0.6$ & $ep=0.8$ & Avg ($ep$) & $ep=0.2$ & $ep=0.4$ & $ep=0.6$ & $ep=0.8$ & Avg ($ep$) \\ \hline
		\textcolor{gray}{GW (Classical)} & \textcolor{gray}{1.0000} & \textcolor{gray}{1.0000} & \textcolor{gray}{1.0000} & \textcolor{gray}{1.0000} & \textcolor{gray}{1.0000} & \textcolor{gray}{1.0000} & \textcolor{gray}{1.0000} & \textcolor{gray}{1.0000} & \textcolor{gray}{1.0000} & \textcolor{gray}{1.0000} & \textcolor{gray}{1.0000} & \textcolor{gray}{1.0000} & \textcolor{gray}{1.0000} & \textcolor{gray}{1.0000} & \textcolor{gray}{1.0000} \\ \hline
		Standard QAOA & 0.8809 & 0.8583 & 0.8977 & 0.8654 & 0.8756 & 0.6395 & 0.9424 & 0.9446 & 0.8807 & 0.8518 & 0.9005 & 0.8923 & 0.8684 & 0.8757 & 0.8842 \\ 
		RBM QAOA & 0.5455 & 0.6689 & 0.6887 & 0.7570 & 0.6650 & 0.5306 & 0.6643 & 0.7398 & 0.7892 & 0.6810 & 0.4756 & 0.6882 & 0.7285 & 0.7684 & 0.6652 \\
		$\mathrm{QAOA}^2$ & 0.6000 & 0.6067 & 0.5333 & 0.5714 & 0.5779 & 0.6735 & 0.6058 & 0.6295 & 0.5956 & 0.6261 & 0.6250 & 0.6250 & 0.6042 & 0.5893 & 0.6109 \\ 
		\textbf{Ours} & 0.8776 & 0.9155 & 0.9397 & 0.9389 & \textbf{0.9179} & 0.8823 & 0.9510 & 0.9402 & 0.8999 & \textbf{0.9184} & 0.9120 & 0.9321 & 0.9004 & 0.9096 & \textbf{0.9135} \\ \hline
		
		\textbf{\multirow{2}*{AR (avg)}} & \multicolumn{5}{c|}{$i=5$} & \multicolumn{5}{c|}{$i=6$} & \multicolumn{5}{c|}{$i=7$} \\ \cline{2-16}
		~ & $ep=0.2$ & $ep=0.4$ & $ep=0.6$ & $ep=0.8$ & Avg ($ep$) & $ep=0.2$ & $ep=0.4$ & $ep=0.6$ & $ep=0.8$ & Avg ($ep$) & $ep=0.2$ & $ep=0.4$ & $ep=0.6$ & $ep=0.8$ & Avg ($ep$) \\ \hline
		\textcolor{gray}{GW (Classical)} & \textcolor{gray}{1.0000} & \textcolor{gray}{0.9522} & \textcolor{gray}{0.9465} & \textcolor{gray}{0.9359} & \textcolor{gray}{0.9587} & \textcolor{gray}{1.0000} & \textcolor{gray}{0.9564} & \textcolor{gray}{0.9375} & \textcolor{gray}{0.9216} & \textcolor{gray}{0.9539} & \textcolor{gray}{1.0000} & \textcolor{gray}{0.9542} & \textcolor{gray}{0.9475} & \textcolor{gray}{0.9221} & \textcolor{gray}{0.956} \\ \hline
		Standard QAOA & 0.7644 & 0.8951 & 0.9603 & 0.7802 & 0.8500 & 0.8666 & 0.8346 & 0.8722 & 0.8979 & 0.8678 & 0.6495 & 0.8330 & 0.8167 & 0.7462 & 0.7614 \\ 
		RBM QAOA & 0.6959 & 0.4738 & 0.6217 & 0.5282 & 0.5799 & 0.5874 & 0.6082 & 0.7676 & 0.7513 & 0.6786 & 0.6813 & 0.5613 & 0.6438 & 0.7375 & 0.6560 \\ 
		$\mathrm{QAOA}^2$ & 0.7095 & 0.7024 & 0.6745 & 0.5479 & 0.6586 & 0.6827 & 0.6611 & 0.6367 & 0.6068 & 0.6468 & 0.5758 & 0.5768 & 0.5839 & 0.5284 & 0.5662 \\
		\textbf{Ours} & 0.8590 & 0.9248 & 0.9734 & 0.8673 & \textbf{0.9061} & 0.9223 & 0.9343 & 0.9386 & 0.9535 & \textbf{0.9372} & 0.8500 & 0.9160 & 0.9012 & 0.8306 & \textbf{0.8745} \\ \hline
		
		\textbf{\multirow{2}*{AR (avg)}} & \multicolumn{5}{c|}{$i=8$} & \multicolumn{5}{c|}{$i=9$} & \multicolumn{5}{c|}{$i=10$} \\ \cline{2-16}
		~ & $ep=0.2$ & $ep=0.4$ & $ep=0.6$ & $ep=0.8$ & Avg ($ep$) & $ep=0.2$ & $ep=0.4$ & $ep=0.6$ & $ep=0.8$ & Avg ($ep$) & $ep=0.2$ & $ep=0.4$ & $ep=0.6$ & $ep=0.8$ & Avg ($ep$) \\ \hline
		\textcolor{gray}{GW (Classical)} & \textcolor{gray}{1.0000} & \textcolor{gray}{0.9571} & \textcolor{gray}{0.9467} & \textcolor{gray}{0.9309} & \textcolor{gray}{0.9587} & \textcolor{gray}{1.0000} & \textcolor{gray}{0.9553} & \textcolor{gray}{0.9469} & \textcolor{gray}{0.9228} & \textcolor{gray}{0.9563} & \textcolor{gray}{1.0000} & \textcolor{gray}{0.9506} & \textcolor{gray}{0.9471} & \textcolor{gray}{0.9303} & \textcolor{gray}{0.9570} \\ \hline
		Standard QAOA & 0.5745 & 0.7737 & 0.8323 & 0.7642 & 0.7362 & 0.5194 & 0.8002 & 0.8105 & 0.7118 & 0.7105 & 0.7017 & 0.7764 & 0.7238 & 0.8102 & 0.7530 \\ 
		RBM QAOA & 0.5326 & 0.6631 & 0.6807 & 0.7477 & 0.6560 & 0.4847 & 0.6461 & 0.7252 & 0.7681 & 0.6560 & 0.4329 & 0.6837 & 0.7208 & 0.7614 & 0.6497 \\ 
		$\mathrm{QAOA}^2$ & 0.5611 & 0.5638 & 0.5067 & 0.5241 & 0.5389 & 0.5533 & 0.5817 & 0.6000 & 0.5654 & 0.5751 & 0.5545 & 0.5852 & 0.5688 & 0.5661 & 0.5687 \\ 
		\textbf{Ours} & 0.8090 & 0.8559 & 0.8912 & 0.8851 & \textbf{0.8603} & 0.7900 & 0.8630 & 0.8679 & 0.8479 & \textbf{0.8422} & 0.8711 & 0.8764 & 0.8390 & 0.8627 & \textbf{0.8623} \\ \hline
		
		\textbf{\multirow{2}*{Time (second)}} & \multicolumn{5}{c|}{$i=5$} & \multicolumn{5}{c|}{$i=6$} & \multicolumn{5}{c|}{$i=7$} \\ \cline{2-16}
		~ & $ep=0.2$ & $ep=0.4$ & $ep=0.6$ & $ep=0.8$ & Avg ($ep$) & $ep=0.2$ & $ep=0.4$ & $ep=0.6$ & $ep=0.8$ & Avg ($ep$) & $ep=0.2$ & $ep=0.4$ & $ep=0.6$ & $ep=0.8$ & Avg ($ep$) \\ \hline
		Standard QAOA & 32.81 & 36.48 & 31.91 & 38.35 & 34.89 & 45.19 & 59.60 & 58.42 & 63.40 & 56.65 & 56.72 & 69.05 & 65.61 & 71.39 & 65.69 \\ 
		RBM QAOA & 75.07 & 80.44 & 68.46 & 77.03 & 75.25 & 141.49 & 194.59 & 181.16 & 166.18 & 170.86 & 171.71 & 264.24 & 313.77 & 288.58 & 259.58 \\ 
		\textbf{Ours} & 10.92 & 12.44 & 11.36 & 13.79 & \textbf{12.13} & 17.87 & 22.47 & 20.61 & 23.36 & \textbf{21.08} & 29.66 & 28.98 & 38.39 & 38.79 & \textbf{33.96} \\ \hline
		
		\textbf{\multirow{2}*{Time (second)}} & \multicolumn{5}{c|}{$i=8$} & \multicolumn{5}{c|}{$i=9$} & \multicolumn{5}{c|}{$i=10$} \\ \cline{2-16}
		~ & $ep$=0.2 & $ep$=0.4 & $ep$=0.6 & $ep$=0.8 & Avg ($ep$) & $ep$=0.2 & $ep$=0.4 & $ep$=0.6 & $ep$=0.8 & Avg ($ep$) & $ep$=0.2 & $ep$=0.4 & $ep$=0.6 & $ep$=0.8 & Avg ($ep$) \\ \hline
		Standard QAOA & 86.48 & 97.25 & 103.65 & 102.09 & 97.37 & 244.41 & 285.41 & 263.30 & 279.31 & 268.11 & 913.76 & 957.67 & 913.62 & 979.61 & 941.17 \\ 
		RBM QAOA & 379.10 & 247.61 & 345.10 & 365.40 & 334.30 & 390.49 & 391.19 & 401.87 & 383.94 & 391.87 & 426.46 & 547.22 & 599.90 & 605.11 & 544.67 \\ 
		\textbf{Ours} & 46.12 & 49.07 & 48.09 & 73.18 & \textbf{54.11} & 121.00 & 146.16 & 150.73 & 139.38 & \textbf{139.32} & 344.01 & 348.39 & 354.67 & 364.00 & \textbf{352.77} \\ \hline
	\end{tabular}
\end{table*}

According to the results in FIG. \ref{fig3}, TABLE \ref{tab2}, and Supplementary \ref{tab6}, Supplementary TABLE \ref{tab7}, we find that in terms of the AR of weighted graphs:

\begin{enumerate}
	\item Compared with the baseline methods, PIL QAOA method shows superior performance on various weighted graphs. For example, for weighted random graphs, the ARs of PIL QAOA are higher by 29.10\% (max) and 23.44\% (avg) than those of $\mathrm{QAOA}^2$. For all weighted graphs, the ARs of PIL QAOA are higher by about 7.98\%~31.42\%(max) and 15.34\%-30.2\%(avg) than baseline methods.
	\item For all weighted graphs with less than 10 nodes, the ARs of PIL QAOA decrease relatively slightly as the number of nodes increases. Meanwhile, in most scenarios, the ARs of PIL QAOA are typically equivalent to and can even surpass that of the GW classical method.
\end{enumerate}

In terms of training time of weighted graphs:

\begin{enumerate}
	\item PIL QAOA has a shorter training time than the baseline methods on various weighted graphs. For weighted graphs, the training time of PIL QAOA is reduced by 78.04\% compared to Standard QAOA and by 74.07\% compared to RBM QAOA.
\end{enumerate}

\begin{figure}[!hbtp]
	\includegraphics[width=0.29\textwidth]{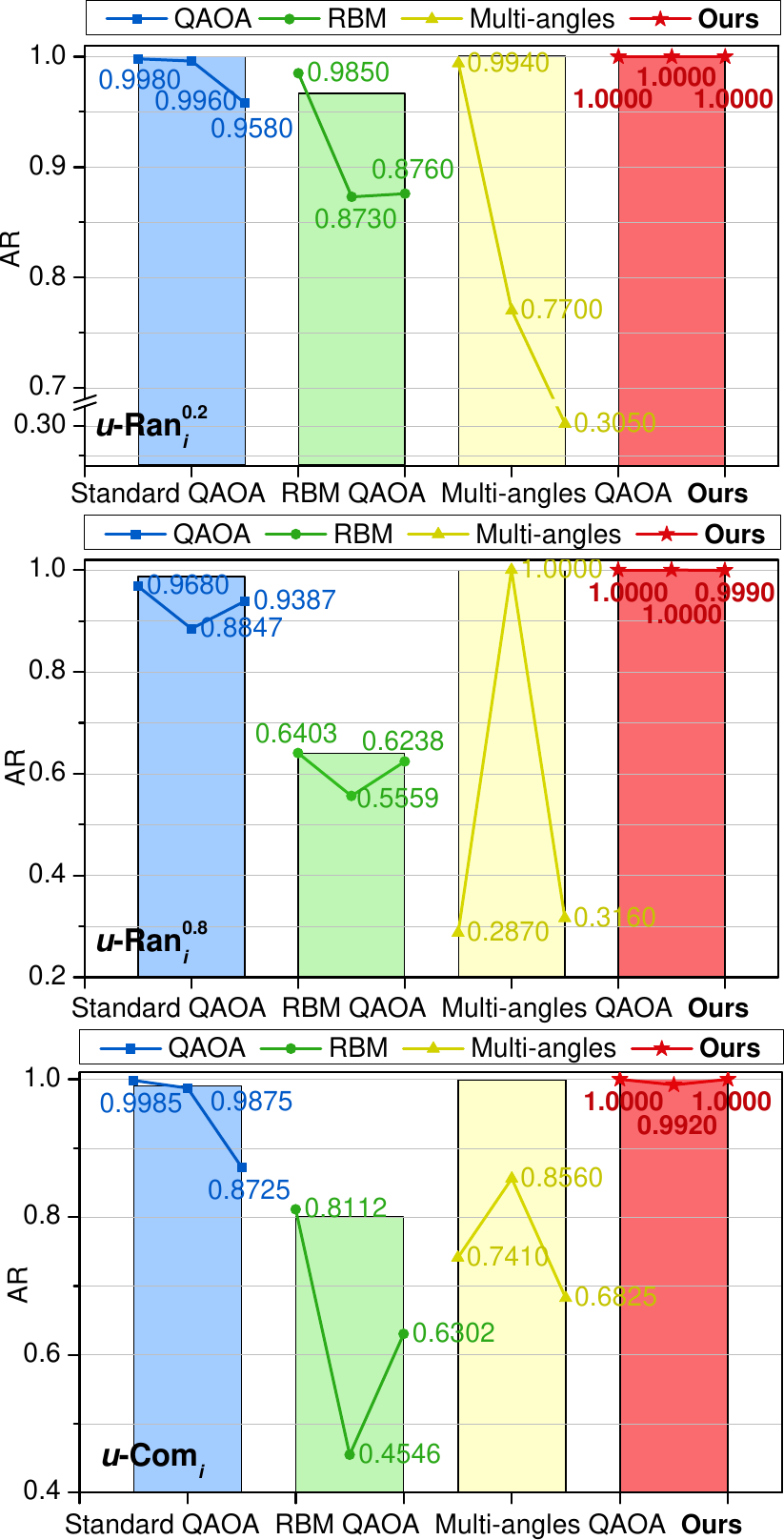}
	\caption{\small{\textbf{Comparison of the anti-forgetting degree of the baseline methods and the proposed method in the unweighted graphs.} The diagram is shown in the lower left corner. Complete graphs and random graphs with edge probabilities of 0.2 and 0.8 are selected for experiments. The column represents the AR of the method on the old graphs (4 nodes), and the broken line shows the AR of applying the optimization parameters of the new graphs (with 6, 8, and 10 nodes) to solve the old graphs.}}
	\label{fig4}
\end{figure} 
\subsubsection{Degree of anti-forgetting}

In the domain of machine learning, the forgetting problem has received widespread attention \cite{60,61}, which causes the model to compromise its performance on old datasets during the optimization process of new datasets. For the MaxCut problems, the performance decline of the model on the old graph after solving the new graph is called forgetting. PIL QAOA exhibits remarkable anti-forgetting abilities and effectively slows down catastrophic amnesia. To demonstrate this point, we compared the anti-forgetting properties of traditional methods and PIL QAOA. The relevant results are shown in FIG. \ref{fig4}.                                               

By observing FIG. \ref{fig4}, we can find that in terms of the degree of anti-forgetting: PIL QAOA shows excellent anti-forgetting performance in different graphs. For example, in unweighted complete graphs, the ARs of PIL QAOA are higher by 25.90\% (6 nodes), 13.60\% (8 nodes), and 31.75\% (10 nodes) than those of Multi-angle QAOA. This result demonstrates that PIL QAOA can effectively retain previously acquired knowledge when optimizing new graphs and will not significantly damage the performance of old graphs. This feature enhances its applicability in the fields of continuous learning and multi-task learning.

\subsubsection{Performance stability}
\begin{figure}[!hbtp]
	\includegraphics[width=0.47\textwidth]{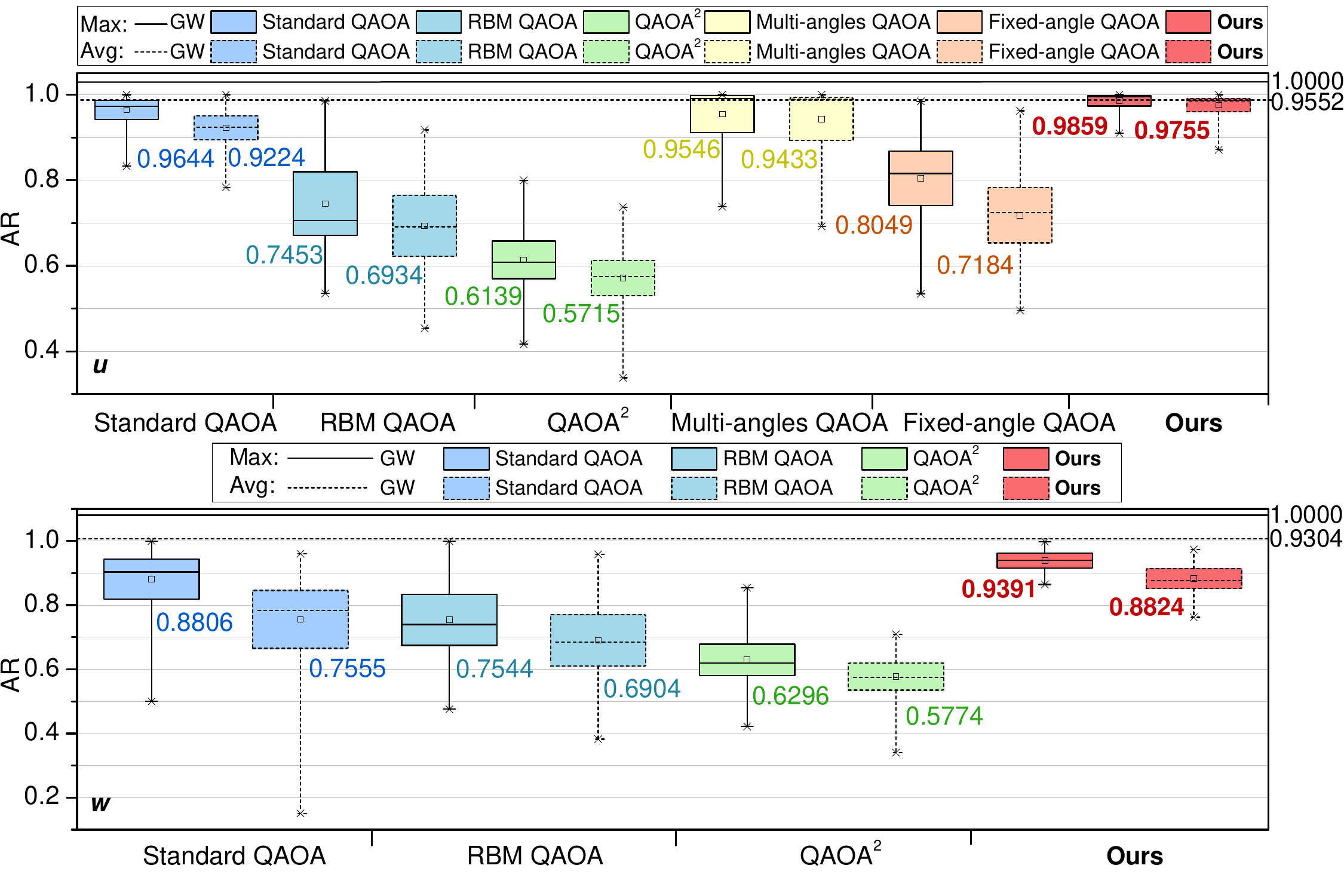}
	\caption{\small{\textbf{Comparison of AR stability between the baseline methods and the proposed method.} The figures above and below represent the results of unweighted graphs and weighted graphs respectively. The rest of the legend information is the same as FIG. \ref{fig2}.}}
	\label{fig5}
\end{figure}

\begin{figure*}[!hbtp]
	\includegraphics[width=0.8\textwidth]{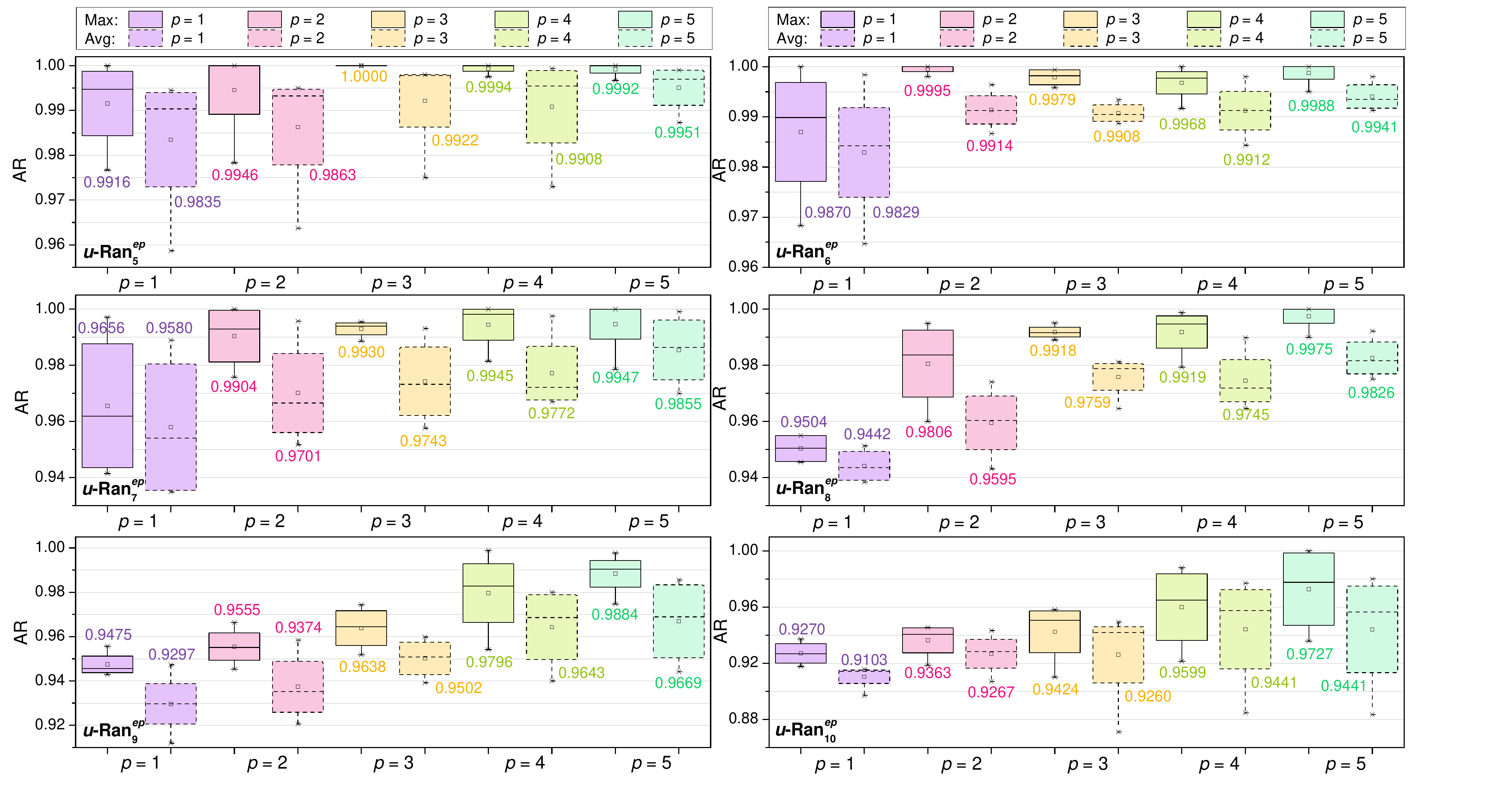}
	\caption{\small{\textbf{Comparison of AR stability between the baseline methods and the proposed method.} The figures above and below represent the results of unweighted graphs and weighted graphs respectively. The rest of the legend information is the same as FIG. \ref{fig2}.}}
	\label{fig6}
\end{figure*}

To ensure that algorithms are reliable and robust in practical applications, methods that show a high level of result stability are generally favored. So this paper conducts a comprehensive comparison of the performance stability of the baseline methods and PIL QAOA on various unweighted and weighted graphs, including random graphs with varying edge probabilities, regular graphs with different degrees of connectivity, and complete graphs. The relevant comparison results are detailed in FIG. \ref{fig5}. In this figure, the smaller the area of the box, the more stable the method is.

According to FIG. \ref{fig5}, PIL QAOA shows superior robustness and reliability compared with the baseline methods on various unweighted and weighted graphs.

\subsubsection{Experiments with different $p$-layers}
In the above simulation experiments, we fix the number of layers (denoted by $p$) to 3. There may be concerns about how to determine a proper $p$ and whether the performance will be enhanced with the $p$ increasing. Taking these into account,we compare the AR of PIL QAOA in unweighted graphs under varying $p$ (1-5). Here are the experimental results of random graphs with high complexity in FIG. \ref{fig6}. The experimental results of other types of graphs are shown in Supplementary information.
 
Based on the different $p$-layer experimental results presented in FIG. \ref{fig6}, Supplementary TABLE \ref{tab8}, Supplementary TABLE \ref{tab9}, and Supplementary TABLE \ref{tab10}, the following observations can be made for the unweighted graph:

\begin{enumerate}
	\item In random graphs with fewer nodes, as $p$ increases, the stability of the AR increases. For example, when $i=5$, the upper and lower limits of the average AR gradually converge with the increase of the number of layers $p$.
	\item For unweighted graphs, as the number of layers increases, the AR generally shows an increasing trend. However, the increase in the number of layers does not result in a significant improvement in the AR. Generally, a deeper level $p$ typically results in extended training durations. Hence, in our experiments, a smaller $p$ is more suitable.
\end{enumerate}

For future research, especially research on larger graphs, the increase in the number of $p$ layers will help improve the approximation ratio of the overall solution, but the degree of improvement does not increase linearly. This inspires us that choosing a larger $p$ requires comprehensive consideration of the effect of the approximate ratio improvement and the cost of extending the training time.

\subsubsection{Outlook}
In this section, we delve into a novel and computationally intricate application: the incremental MaxCut problem. We further explore the inherent advantages of PIL QAOA in addressing such issues.

The incremental MaxCut is an ever-increasing scale MaxCut problem, commonly observed in real-world scenarios like the increase in the number of users in social networks \cite{62}. As the size and topology of the graph grow, so does the complexity of deploying QAOA. Traditional QAOA methods often choose complete re-optimization when faced with the ever-expanding MaxCut. Admittedly, this strategy poses computational power challenges including the lengthy period of training time and the intensive computational resources. For example, when a 4-node MaxCut problem scales with new $n$ nodes and associated edges, standard techniques tend to re-optimize the entire graph that now contains $4+n$ nodes, thus amplifying the computational requirements.

Excitingly, we discover that our method can elegantly address these challenges. PIL QAOA shows obvious superiority in this regard and exhibits excellent performance when dealing with incremental problems. PIL QAOA can effectively learn optimized parameters for smaller graphs and apply them to parameter initialization for incremental graphs. Next, by fine-tuning these parameters, the performance of solving the MaxCut problem on incremental graphs is significantly improved. This method avoids unnecessary training time overhead and opens a new door for practical applications of optimization problems.

\section{Discussion}
In this paper, a novel QAOA, called PIL QAOA, is proposed from a unique perspective to get a superior performance. We proactively transform the training way into an incremental learning approach. We transform the MaxCut problem solution for large graphs into a combination of small graph solutions and incremental learning step-by-step solutions. Although the training is divided into multiple phases, training on relatively simple subgraphs points the way to parameter initialization for large graph problems, thus achieving better performance. Experiments on the MaxCut-Sandbox datasets show that PIL QAOA exhibits superior AR, stability, and anti-forgetting, and significantly shortens the training time. In addition, we also explore the impact of different choices of $p$ in the AR, providing some reference for the design of future methods.

Notably, PIL QAOA hasn't been tested on large-scale problems due to the limitation of simulating quantum computation via classical computers. The performance of our method on large-scale graphs remains an area that requires further validation in the future. Furthermore, besides the MaxCut problem investigated in this paper, exploring the application of the presented incremental method on other optimization problems is significant as well. Although PIL QAOA is a quantum algorithm, we hope the idea of converting problems into an incremental learning approach can also be applied in the future to optimize the solution process of classical algorithms, offering a fresh perspective for problem resolution.

\section{Method}
\subsection{Early break mechanism}
In existing QAOA methods, the classic optimizer often terminates when the gradient convergence, resulting in unnecessary time overhead. We believe that the optimization parameters reused from the previous step allow us to reduce the solution scope of the incremental graphs. To enhance the efficiency of our parameters reuse strategy and decrease redundant optimization, we designed the early break mechanism in the incremental learning phases. This mechanism consists of two modules, i.e. the randomly solved module and the comparison module. The randomly solved module requires solving multiple times randomly and choosing the best solution as the output. The comparison module will compare the output of the randomly solved module to the current optimized solution. Training is stopped early if the current solution is greater or equal to the random solution. On the contrary, optimization will persist. Algorithm \ref{Alga} shows the detailed steps of the proposed mechanism.

\begin{algorithm}[H]
	\caption{Early break mechanism of PIL QAOA.} 
	\label{Alga}
	\begin{algorithmic}[1]
		\renewcommand{\algorithmicrequire}{\textbf{Input:}}
		\Require Incremental graph $G'$, current parameters $\psi$.
		\State Initialize the early break value of Boolean variable “stop-optimize” to False, $C(z)$ and $C'(z)$ to negative infinity;
		\For{$i=1,2$, $\dots$, $k$}:
			\State Randomly divide the incremental graph $G'$ into node sets $V_+$ and $V_-$ multiple times to obtain the corresponding divided bit strings $z_i$;
			\State Calculate the cut value $C(z_i)$ corresponding to $z_i$;
		\EndFor
		\State Sort $C(z_i)$ to get the maximum value $C(z)_{max}$;
		\State From the current parameters $\psi$, $C'(z)$ of the incremental graph $y$ is obtained;
		\If{$C'(z)\geq C(z)_{max}$}:
		 	\State Set the stop-optimize value to True;
		\EndIf
		\If{the stop-optimize value is True}:
		    \State Jump out of the optimization process of the current incremental graph $G'$ and enter the optimization process of the next incremental graph or target graph $G''$;
		\Else
			\State Optimization of the currently incremental graph $G'$ is continued;
		\EndIf
		\State \textbf{Repeat the above steps until entering the optimization process of the target graphs.}
	\end{algorithmic} 
\end{algorithm}

\subsection{Classic optimizer}
In PIL QAOA, a classic optimizer is employed to optimize the graph's parameters. Gradient descent algorithms can be used such as COBYLA (Constrained Optimization BY Linear Approximation) \cite{63,64,65}, SLSQP (Sequential Least Squares Programming) \cite{66}, BFGS (Broyden-Fletcher-Goldfarb-Shanno) \cite{67}, and BOBYQA (Bound Optimization BY Quadratic Approximation) \cite{68,69} to achieve the convergence of cost values. In the experimental part, we choose the COBYLA algorithm, which is simple and has a good optimization effect.

\subsection{Experiments Settings}
This section provides a comprehensive account of the experimental settings, including dataset construction, and evaluation metrics.

\subsubsection{Dataset construction}
To assess the effectiveness of PIL QAOA in solving MaxCut problems, we construct a novel and comprehensive dataset called "MaxCut-Sandbox" for QAOA. MaxCut-Sandbox consists of six graph types: unweighted: random, regular, and complete graphs; weighted: random, regular, and complete graphs. For the experiment, we set the number of nodes in the base graph to 4 and chose a range of 5 to 10 nodes for the target graph to observe AR. (Considering the performance constraints of computational simulation, the upper limit of the number of nodes is also selected to be about 10, which is selected in most experimental parts of existing work.) One node with corresponding edges is added to the subgraph in each phase. The specific information of the MaxCut-Sandbox is shown in TABLE \ref{tab3}.

\begin{table}[!h]
	\begin{center}
		\centering
		\footnotesize
		\caption{Specific information of the MaxCut-Sandbox.}\label{tab3}
		\begin{tabular}{c|ccc}
			\hline
			\multirow{2}*{Random} & Number of nodes  & Probability of edges & Symbol \\ \cline{2-4}
			~ & 5, 6, 7, 8, 9, 10 & 0.2, 0.4, 0.6, 0.8 & $x\mathrm{-Ran}_i^{ep}$ \\ \hline
			\multirow{4}*{Regular} & Number of nodes  & Degree & Symbol \\ \cline{2-4}
			~ & 5, 6, 7, 8, 9, 10 & 2 & ~ \\ 
			~ & 6, 8, 10 & 3 & $x\mathrm{-Reg}_i^{d}$ \\ 
			~ & 5, 6, 7, 8, 9, 10 & 4 & ~ \\ \hline
			\multirow{2}*{Complete} & \multicolumn{2}{c}{Number of nodes} & Symbol \\ \cline{2-4}
			~ & \multicolumn{2}{c}{5, 6, 7, 8, 9, 10} & $x\mathrm{-Com}_i$ \\ \hline
		\end{tabular}
	\end{center}
	\begin{tablenotes}
		\item[] Footnote:  $x$ represents whether there is authority, $x={u,w}$, when $x$ is $u$, it is an unweighted graph; when $x$ is $w$, it is a weighted graph. $ep$ represents the probability of generating an edge, with $ep$=\{0.2, 0.4, 0.6, 0.8\}. $i$ represents the number of nodes, $i=\{5,6,7,8,9,10\}$. $d$ represents the degree, with $d=\{2,3,4\}$. For example, $u\mathrm{-Ran}_6^{0.4}$ represents an unweighted random graph with 6 nodes with edge probability 0.4; $w\mathrm{-Reg}_9^{4}$ represents a weighted regular graph with 9 nodes of degree 4.
	\end{tablenotes}
\end{table}

\subsubsection{Evaluation metrics}

To evaluate the performance of the methods, we use AR as one of the key evaluation metrics \cite{39,40,41,42}. This index represents the ratio between the approximate solution $C_A$ obtained by the method and the actual optimal solution $C^*$ to the problem. Its formula is shown below:
\begin{equation}
	\mathrm{AR}=\frac{C_A}{C^*},
\end{equation}

The range of AR values is $[-1,1]$. The greater the absolute value of the AR value, the better the performance will be.

To eliminate the randomness of the optimization process as much as possible, we repeat each solution process 10 times, and calculate the maximum value and average value of these 10 occurrences. 

Furthermore, we utilize T to evaluate the training time of the methods. Its formula is shown below:
\begin{equation}
\mathrm{T}=T_{end}-T_{start},
\end{equation}

\noindent where, $T_{end}$ and $T_{start}$ represent the timestamp of ending training and starting training.

\newpage

\section*{Acknowledgements}
This work is supported by National Natural Science Foundation of China (Grant Nos. 62372048, 62371069, 62272056)

\clearpage
\newpage
\appendix

\onecolumngrid
 
\begin{center}
\large{\textbf{Supplementary Information: ``Proactively incremental-learning QAOA''}}	
\end{center}

\begin{table}[!ht]
	\centering
	\caption{\small{\textbf{Comparison of AR and Time between the baseline methods and the proposed method in unweighted random graphs.}}}\label{tab4}
	\tiny
	\begin{tabular}{|c|cccc|c|cccc|c|cccc|c|}
		\hline
		\textbf{\multirow{2}*{AR (max)}} & \multicolumn{5}{c|}{$i=5$} & \multicolumn{5}{c|}{$i=6$} & \multicolumn{5}{c|}{$i=7$} \\ \cline{2-16}
		~ & $ep=0.2$ & $ep=0.4$ & $ep=0.6$ & $ep=0.8$ & Avg ($ep$)& $ep=0.2$ & $ep=0.4$ & $ep=0.6$ & $ep=0.8$ & Avg ($ep$) & $ep=0.2$ & $ep=0.4$ & $ep=0.6$ & $ep=0.8$ & Avg ($ep$) \\ \hline
		\textcolor{gray}{GW (Classical)} & \textcolor{gray}{1.0000} & \textcolor{gray}{1.0000} & \textcolor{gray}{1.0000} & \textcolor{gray}{1.0000} & \textcolor{gray}{1.0000} & \textcolor{gray}{1.0000} & \textcolor{gray}{1.0000} & \textcolor{gray}{1.0000} & \textcolor{gray}{1.0000} & \textcolor{gray}{1.0000} & \textcolor{gray}{1.0000} & \textcolor{gray}{1.0000} & \textcolor{gray}{1.0000} & \textcolor{gray}{1.0000} & \textcolor{gray}{1.0000} \\ \hline
		Standard QAOA & 0.9725 & 0.9940 & 0.9950 & 0.9250 & 0.9716 & 0.9700 & 0.9783 & 0.9860 & 0.9912 & 0.9814 & 1.0000 & 0.9200 & 0.9689 & 0.9645 & 0.9634 \\ 
		RBM QAOA & 0.8375 & 0.6845 & 0.7831 & 0.7038 & 0.7522 & 0.7506 & 0.6575 & 0.8157 & 0.7469 & 0.7427 & 0.6963 & 0.5672 & 0.6541 & 0.7822 & 0.6749 \\ 
		$\mathrm{QAOA}^2$ & 0.7500 & 0.7250 & 0.7500 & 0.5938 & 0.7047 & 0.6625 & 0.7051 & 0.6500 & 0.6484 & 0.6665 & 0.6000 & 0.6429 & 0.5764 & 0.5284 & 0.5869 \\ 
		Multi-angle QAOA & 0.9999 & 1.0000 & 1.0000 & 1.0000 & 1.0000 & 0.9998 & 0.9998 & 0.9999 & 0.9999 & 0.9999 & 0.9947 & 0.9969 & 0.9942 & 0.9987 & 0.9961 \\ 
		Fixed-angle QAOA & 0.7629 & 0.6712 & 0.8181 & 0.8969 & 0.7873 & 0.6854 & 0.6621 & 0.8157 & 0.8157 & 0.7447 & 0.5342 & 0.7505 & 0.9289 & 0.8012 & 0.7537 \\ 
		\textbf{Ours} & 1.0000 &1.0000 & 1.0000 & 1.0000 & \textbf{1.0000} & 0.9970 & 0.9994 & 0.9958 & 0.9994 & \textbf{0.9979} & 0.9947 & 0.9886 & 0.9931 & 0.9955 & \textbf{0.9930} \\ \hline
		
		\textbf{\multirow{2}*{AR (max)}} & \multicolumn{5}{c|}{$i=8$} & \multicolumn{5}{c|}{$i=9$} & \multicolumn{5}{c|}{$i=10$} \\ \cline{2-16}
		~ & $ep=0.2$ & $ep=0.4$ & $ep=0.6$ & $ep=0.8$ & Avg ($ep$)& $ep=0.2$ & $ep=0.4$ & $ep=0.6$ & $ep=0.8$ & Avg ($ep$) & $ep=0.2$ & $ep=0.4$ & $ep=0.6$ & $ep=0.8$ & Avg ($ep$) \\ \hline
		\textcolor{gray}{GW (Classical)} & \textcolor{gray}{1.0000} & \textcolor{gray}{1.0000} & \textcolor{gray}{1.0000} & \textcolor{gray}{1.0000} & \textcolor{gray}{1.0000} & \textcolor{gray}{1.0000} & \textcolor{gray}{1.0000} & \textcolor{gray}{1.0000} & \textcolor{gray}{1.0000} & \textcolor{gray}{1.0000} & \textcolor{gray}{1.0000} & \textcolor{gray}{1.0000} & \textcolor{gray}{1.0000} & \textcolor{gray}{1.0000} & \textcolor{gray}{1.0000} \\ \hline
		Standard QAOA & 0.8900 & 0.9470 & 0.9391 & 0.9743 & 0.9376 & 0.8856 & 0.9500 & 0.9464 & 0.9629 & 0.9362 & 0.8327 & 0.9231 & 0.9289 & 0.9124 & 0.8993 \\ 
		RBM QAOA & 0.5693 & 0.6791 & 0.6743 & 0.7302 & 0.6632 & 0.5353 & 0.6631 & 0.7064 & 0.7634 & 0.6670 & 0.5420 & 0.6859 & 0.7266 & 0.7594 & 0.6785 \\ 
		$\mathrm{QAOA}^2$ & 0.6083 & 0.5750 & 0.6000 & 0.5491 & 0.5831 & 0.6319 & 0.6154 & 0.6384 & 0.5956 & 0.6203 & 0.6477 & 0.5977 & 0.5868 & 0.5685 & 0.6002 \\ 
		Multi-angle QAOA & 0.9768 & 0.9825 & 0.9829 & 0.9757 & 0.9795 & 0.8684 & 0.8990 & 0.9526 & 0.9265 & 0.9116 & 0.7830 & 0.8215 & 0.8603 & 0.8606 & 0.8314 \\ 
		Fixed-angle QAOA & 0.6274 & 0.7644 & 0.7816 & 0.8431 & 0.7541 & 0.5573 & 0.7265 & 0.8065 & 0.7573 & 0.7119 & 0.5975 & 0.7183 & 0.7731 & 0.7865 & 0.7188 \\ 
		\textbf{Ours} & 0.9951 & 0.9890 & 0.9920 & 0.9912 & \textbf{0.9918} & 0.9744 & 0.9601 & 0.9689 & 0.9518 & \textbf{0.9638} & 0.9582 & 0.9562 & 0.9450 & 0.9100 & \textbf{0.9424} \\ \hline
		
		\textbf{\multirow{2}*{AR (avg)}} & \multicolumn{5}{c|}{$i=5$} & \multicolumn{5}{c|}{$i=6$} & \multicolumn{5}{c|}{$i=7$} \\ \cline{2-16}
		~ & $ep=0.2$ & $ep=0.4$ & $ep=0.6$ & $ep=0.8$ & Avg ($ep$)& $ep=0.2$ & $ep=0.4$ & $ep=0.6$ & $ep=0.8$ & Avg ($ep$) & $ep=0.2$ & $ep=0.4$ & $ep=0.6$ & $ep=0.8$ & Avg ($ep$) \\ \hline
		\textcolor{gray}{GW (Classical)} & \textcolor{gray}{0.9273} & \textcolor{gray}{0.9450} & \textcolor{gray}{0.9544} & \textcolor{gray}{0.9457} & \textcolor{gray}{0.9431} & \textcolor{gray}{0.9218} & \textcolor{gray}{0.9338} & \textcolor{gray}{0.9578} & \textcolor{gray}{0.9429} & \textcolor{gray}{0.9390} & \textcolor{gray}{0.9273} & \textcolor{gray}{0.9250} & \textcolor{gray}{0.9500} & \textcolor{gray}{0.9448} & \textcolor{gray}{0.9368} \\ \hline
		Standard QAOA & 0.9260 & 0.9600 & 0.9560 & 0.8993 & 0.9353 & 0.9112 & 0.9290 & 0.9664 & 0.9502 & 0.9392 & 0.9037 & 0.9034 & 0.9240 & 0.9495 & 0.9202 \\ 
		RBM QAOA & 0.8155 & 0.5850 & 0.6915 & 0.6507 & 0.6857 & 0.7014 & 0.6281 & 0.7472 & 0.6952 & 0.6930 & 0.6750 & 0.5574 & 0.6505 & 0.7755 & 0.6646 \\ 
		$\mathrm{QAOA}^2$ & 0.6898 & 0.6568 & 0.6728 & 0.5718 & 0.6478 & 0.6280 & 0.6410 & 0.6018 & 0.5963 & 0.6168 & 0.5733 & 0.5788 & 0.5284 & 0.4970 & 0.5444 \\
		Multi-angle QAOA & 0.9999 & 0.9999 & 0.9999 & 0.9999 & 0.9999 & 0.9993 & 0.9994 & 0.9999 & 0.9995 & 0.9995 & 0.9941 & 0.9950 & 0.9928 & 0.9976 & 0.9949 \\ 
		Fixed-angle QAOA & 0.6526 & 0.6525 & 0.7176 & 0.7733 & 0.6990 & 0.6072 & 0.6113 & 0.7796 & 0.7796 & 0.6944 & 0.4954 & 0.6852 & 0.8462 & 0.7459 & 0.6932 \\ 
		\textbf{Ours} & 0.9980 & 0.9976 & 0.9980 & 0.9750 & \textbf{0.9922} & 0.9934 & 0.9887 & 0.9895 & 0.9914 & \textbf{0.9908} & 0.9931 & 0.9666 & 0.9799 & 0.9576 & \textbf{0.9743} \\ \hline
		
		\textbf{\multirow{2}*{AR (avg)}} & \multicolumn{5}{c|}{$i=8$} & \multicolumn{5}{c|}{$i=9$} & \multicolumn{5}{c|}{$i=10$} \\ \cline{2-16}
		~ & $ep=0.2$ & $ep=0.4$ & $ep=0.6$ & $ep=0.8$ & Avg ($ep$)& $ep=0.2$ & $ep=0.4$ & $ep=0.6$ & $ep=0.8$ & Avg ($ep$) & $ep=0.2$ & $ep=0.4$ & $ep=0.6$ & $ep=0.8$ & Avg ($ep$) \\ \hline
		\textcolor{gray}{GW (Classical)} & \textcolor{gray}{0.9364} & \textcolor{gray}{0.9225} & \textcolor{gray}{0.9511} & \textcolor{gray}{0.9429} & \textcolor{gray}{0.9382} & \textcolor{gray}{0.9382} & \textcolor{gray}{0.9338} & \textcolor{gray}{0.9533} & \textcolor{gray}{0.9467} & \textcolor{gray}{0.9430} & \textcolor{gray}{0.9291} & \textcolor{gray}{0.9100} & \textcolor{gray}{0.9539} & \textcolor{gray}{0.9410} & \textcolor{gray}{0.9335} \\ \hline
		Standard QAOA & 0.8444 & 0.9130 & 0.9127 & 0.9474 & 0.9044 & 0.8293 & 0.8865 & 0.9304 & 0.8747 & 0.8802 & 0.7834 & 0.8901 & 0.8671 & 0.8521 & 0.8482 \\ 
		RBM QAOA & 0.5475 & 0.6581 & 0.6506 & 0.7127 & 0.6422 & 0.5113 & 0.6563 & 0.7024 & 0.7625 & 0.6582 & 0.4808 & 0.6721 & 0.7230 & 0.7559 & 0.6580 \\ 
		$\mathrm{QAOA}^2$ & 0.5480 & 0.5287 & 0.5387 & 0.5164 & 0.5330 & 0.5750 & 0.5904 & 0.6018 & 0.5699 & 0.5843 & 0.5955 & 0.5711 & 0.5715 & 0.5589 & 0.5743 \\ 
		Multi-angle QAOA & 0.9607 & 0.9665 & 0.9674 & 0.9656 & 0.9651 & 0.8187 & 0.8782 & 0.9174 & 0.9127 & 0.8818 & 0.7411 & 0.8113 & 0.8325 & 0.8496 & 0.8086 \\ 
		Fixed-angle QAOA & 0.5306 & 0.7095 & 0.7512 & 0.8201 & 0.7028 & 0.5196 & 0.7072 & 0.7479 & 0.7400 & 0.6787 & 0.5666 & 0.6974 & 0.7442 & 0.7696 & 0.6944 \\ 
		\textbf{Ours} & 0.9812 & 0.9646 & 0.9776 & 0.9801 & \textbf{0.9759} & 0.9598 & 0.9466 & 0.9551 & 0.9392 & \textbf{0.9502} & 0.9409 & 0.9492 & 0.9428 & 0.8711 & \textbf{0.9260} \\ \hline
		
		\textbf{\multirow{2}*{Time (second)}} & \multicolumn{5}{c|}{$i=5$} &\multicolumn{5}{c|}{$i=6$} & \multicolumn{5}{c|}{$i=7$} \\ \cline{2-16}
		~ & $ep=0.2$ & $ep=0.4$ & $ep=0.6$ & $ep=0.8$ & Avg ($ep$)& $ep=0.2$ & $ep=0.4$ & $ep=0.6$ & $ep=0.8$ & Avg ($ep$) & $ep=0.2$ & $ep=0.4$ & $ep=0.6$ & $ep=0.8$ & Avg ($ep$) \\ \hline
		Standard QAOA & 27.75 & 40.23 & 37.45 & 46.44 & 37.97 & 45.75 & 55.04 & 68.30 & 68.95 & 59.51 & 59.71 & 60.63 & 86.64 & 90.99 & 74.49 \\ 
		RBM QAOA & 76.62 & 83.18 & 56.92 & 89.98 & 76.67 & 82.94 & 91.00 & 61.89 & 147.27 & 95.77 & 97.93 & 72.08 & 95.39 & 152.22 & 104.40 \\ 
		Multi-angle QAOA & 18.15 & 10.11 & 10.69 & 11.14 & 12.52 & 20.89 & 21.84 & 20.29 & 20.81 & 20.96 & 52.15 & 47.27 & 50.60 & 40.92 & 47.73 \\ 
		\textbf{Ours} & 11.02 & 11.65 & 13.85 & 13.91 & \textbf{12.61} & 18.55 & 18.43 & 21.72 & 22.93 & \textbf{20.41} & 26.32 & 24.11 & 32.62 & 33.68 & \textbf{29.18} \\ \hline
		
		\textbf{\multirow{2}*{Time (second)}} & \multicolumn{5}{c|}{$i=8$} &\multicolumn{5}{c|}{$i=9$} & \multicolumn{5}{c|}{$i=10$} \\ \cline{2-16}
		~ & $ep=0.2$ & $ep=0.4$ & $ep=0.6$ & $ep=0.8$ & Avg ($ep$)& $ep=0.2$ & $ep=0.4$ & $ep=0.6$ & $ep=0.8$ & Avg ($ep$) & $ep=0.2$ & $ep=0.4$ & $ep=0.6$ & $ep=0.8$ & Avg ($ep$) \\ \hline
		Standard QAOA & 86.56 & 104.21 & 145.00 & 126.04 & 115.45 & 256.92 & 275.41 & 290.15 & 294.53 & 279.25 & 864.14 & 908.01 & 926.57 & 934.61 & 908.33 \\ 
		RBM QAOA & 168.86 & 154.49 & 185.60 & 162.19 & 167.78 & 182.28 & 177.99 & 220.38 & 238.94 & 204.90 & 290.95 & 341.15 & 407.57 & 496.76 & 384.11 \\ 
		Multi-angle QAOA & 101.70 & 88.56 & 97.59 & 84.25 & 93.02 & 198.66 & 201.26 & 193.38 & 211.27 & 201.14 & 369.17 & 362.54 & 384.03 & 401.39 & 379.28 \\ 
		\textbf{Ours} & 40.33 & 39.27 & 48.03 & 53.08 & \textbf{45.18} & 94.88 & 92.39 & 112.28 & 117.96 & \textbf{104.38} & 284.62 & 268.37 & 303.01 & 349.61 & \textbf{301.40} \\ \hline
	\end{tabular}
 	\begin{tablenotes}
		\item[] Note: Each experiment was repeated ten times for every graph, and the results were averaged to provide. \textcolor{gray}{Gray} represents the performance of the classical GW method. \textbf{Bold} represents the best AR value of these QAOA methods.
	\end{tablenotes}
\end{table}

\begin{table}[!ht]
	\centering
	\caption{\small{\textbf{Comparison of AR between the baseline methods and the proposed method in unweighted complete graphs.} }}\label{tab5}
	\tiny
	\begin{tabular}{|c|cccccc|c|}
		\hline
		\textbf{AR (max)} & $i=5$ & $i=6$ & $i=7$ & $i=8$ & $i=9$ & $i=10$ & Avg ($i$) \\ \hline
		\textcolor{gray}{GW (Classic)} & \textcolor{gray}{1.0000} & \textcolor{gray}{1.0000} & \textcolor{gray}{1.0000} & \textcolor{gray}{1.0000} & \textcolor{gray}{1.0000} & \textcolor{gray}{1.0000} & \textcolor{gray}{1.0000} \\ \hline
		Standard QAOA & 1.0000 & 1.0000 & 0.9983 & 0.9775 & 0.9980 & 0.9784 & 0.9920 \\ 
		RBM QAOA & 0.9856 & 0.9247 & 0.9365 & 0.8800 & 0.9422 & 0.8859 & 0.9258 \\ 
		$\mathrm{QAOA}^2$ & 0.6111 & 0.4907 & 0.4167 & 0.4375 & 0.5844 & 0.5875 & 0.5213 \\ 
		Multi-angle QAOA & 1.0000 & 0.9998 & 0.9997 & 0.9907 & 0.9903 & 0.9687 & 0.9915 \\ 
		Fixed-angle QAOA & 0.8969 & 0.8452 & 0.8370 & 0.8049 & 0.8008 & 0.8265 & 0.8352 \\ 
		\textbf{Ours} & 1.0000 & 0.9998 & 0.9997 & 0.9981 & 0.9997 & 0.9796 & \textbf{0.9962} \\ \hline
		
		\textbf{AR (avg)} & $i=5$ & $i=6$ & $i=7$ & $i=8$ & $i=9$ & $i=10$ & Avg ($i$) \\ \hline
		\textcolor{gray}{GW (Classic)} & \textcolor{gray}{0.9667} & \textcolor{gray}{0.9400} & \textcolor{gray}{0.9633} & \textcolor{gray}{0.9625} & \textcolor{gray}{0.9660} & \textcolor{gray}{0.9688} & \textcolor{gray}{0.9612} \\ \hline
		Standard QAOA & 0.9773 & 0.9622 & 0.9540 & 0.9204 & 0.9190 & 0.9118 & 0.9408 \\ 
		RBM QAOA & 0.9178 & 0.8761 & 0.8659 & 0.8381 & 0.8991 & 0.8173 & 0.8690 \\ 
		$\mathrm{QAOA}^2$ & 0.5773 & 0.4699 & 0.3892 & 0.3380 & 0.5762 & 0.5645 & 0.4859 \\ 
		Multi-angle QAOA & 1.0000 & 0.9998 & 0.9994 & 0.9888 & 0.9853 & 0.9625 & 0.9893 \\ 
		Fixed-angle QAOA & 0.7733 & 0.7614 & 0.5556 & 0.6797 & 0.7217 & 0.7247 & 0.7027 \\ 
		\textbf{Ours} & 1.0000 & 0.9996 & 0.9982 & 0.9930 & 0.9859 & 0.9679 & \textbf{0.9908} \\ \hline
		
		\textbf{Time (second)} & $i=5$ & $i=6$ & $i=7$ & $i=8$ & $i=9$ & $i=10$ & Avg ($i$) \\ \hline
		Standard QAOA & 47.07 & 76.22 & 91.77 & 125.24 & 329.81 & 972.97 & 273.84 \\ 
		RBM QAOA & 50.64 & 74.33 & 78.33 & 128.26 & 145.98 & 301.54 & 129.85 \\ 
		Multi-angle QAOA & 9.49 & 24.94 & 40.58 & 87.47 & 188.47 & 398.30 & 124.87 \\ 
		\textbf{Ours} & 12.96 & 27.47 & 38.71 & 59.13 & 116.47 & 304.58 & \textbf{93.22} \\ \hline
	\end{tabular}
        \begin{tablenotes}
		\item[] Note: Each experiment was repeated ten times for every graph, and the results were averaged to provide. \textcolor{gray}{Gray} represents the performance of the classical GW method. \textbf{Bold} represents the best AR value of these QAOA methods.
	\end{tablenotes}
\end{table}

\begin{table}[!ht]
	\centering
	\caption{\small{\textbf{Comparison of AR between the baseline methods and the proposed method in weighted regular graphs.} }}\label{tab6}
	\tiny
	\begin{tabular}{|c|cccccc|c|ccc|c|cccccc|c|}
		\hline
		\textbf{\multirow{2}*{AR (max)}} & \multicolumn{7}{c|}{$d=2$} & \multicolumn{4}{c|}{$d=3$} & \multicolumn{7}{c|}{$d=4$} \\ \cline{2-19}
		\textbf{} & $i=5$ & $i=6$ & $i=7$ & $i=8$ & $i=9$ & $i=10$ & Avg ($i$) & $i=6$ & $i=8$ & $i=10$ & Avg ($i$) & $i=5$ & $i=6$ & $i=7$ & $i=8$ & $i=9$ & $i=10$ & Avg ($i$) \\ \hline
		\textcolor{gray}{GW (Classic)} & \textcolor{gray}{1.0000} & \textcolor{gray}{1.0000} & \textcolor{gray}{1.0000} & \textcolor{gray}{1.0000} & \textcolor{gray}{1.0000} & \textcolor{gray}{1.0000} & \textcolor{gray}{1.0000}  & \textcolor{gray}{1.0000} & \textcolor{gray}{1.0000} & \textcolor{gray}{1.0000} & \textcolor{gray}{1.0000} & \textcolor{gray}{1.0000} & \textcolor{gray}{1.0000} & \textcolor{gray}{1.0000}  & \textcolor{gray}{1.0000} & \textcolor{gray}{1.0000} & \textcolor{gray}{1.0000} & \textcolor{gray}{1.0000}  \\ \hline
		Standard QAOA & 0.9059  & 0.9695  & 0.8186  & 0.9448  & 0.9786  & 0.7507  & 0.8947  & 1.0000  & 0.9117  & 0.8258  & 0.9125  & 0.9906  & 0.9435  & 0.4998  & 0.8079  & 0.7593  & 0.6103  & 0.7686  \\
		RBM QAOA & 0.9152  & 0.9996  & 0.9470  & 0.8999  & 0.6844  & 0.6659  & 0.8520  & 0.7568  & 0.6418  & 0.5954  & 0.6646  & 0.9801  & 0.9232  & 0.8714  & 0.7081  & 0.6422  & 0.6216  & 0.7911  \\
		$\mathrm{QAOA}^2$ & 0.8542  & 0.6818  & 0.7333  & 0.7048  & 0.6172  & 0.6667  & 0.7097  & 0.5438  & 0.6172  & 0.6312  & 0.5974  & 0.5729  & 0.6339  & 0.6295  & 0.5812  & 0.6202  & 0.6523  & 0.6150  \\ 
		\textbf{Ours} & 0.9378  & 0.9981  & 0.9712  & 0.9637  & 0.9578  & 0.8644  & \textbf{0.9488}  & 0.9912  & 0.9743  & 0.8998  & \textbf{0.9551}  & 0.9968  & 0.9799  & 0.9231  & 0.8938  & 0.8927  & 0.8913  & \textbf{0.9296}  \\ \hline
		
		\textbf{\multirow{2}*{AR (avg)}} & \multicolumn{7}{c|}{$d=2$} & \multicolumn{4}{c|}{$d=3$} & \multicolumn{7}{c|}{$d=4$} \\ \cline{2-19}
		\textbf{} & $i=5$ & $i=6$ & $i=7$ & $i=8$ & $i=9$ & $i=10$ & Avg ($i$) & $i=6$ & $i=8$ & $i=10$ & Avg ($i$) & $i=5$ & $i=6$ & $i=7$ & $i=8$ & $i=9$ & $i=10$ & Avg ($i$) \\ \hline
		\textcolor{gray}{GW (Classic)} & \textcolor{gray}{1.0000}  & \textcolor{gray}{0.9597}  & \textcolor{gray}{0.8729}  & \textcolor{gray}{1.0000}  & \textcolor{gray}{0.7661}  & \textcolor{gray}{0.8935}  & \textcolor{gray}{0.9154}  & \textcolor{gray}{1.0000}  & \textcolor{gray}{0.8972}  & \textcolor{gray}{0.8905}  & \textcolor{gray}{0.9292}  & \textcolor{gray}{1.0000}  & \textcolor{gray}{0.9999}  & \textcolor{gray}{0.5549}  & \textcolor{gray}{0.7605}  & \textcolor{gray}{0.7986}  & \textcolor{gray}{0.8545}  & \textcolor{gray}{0.8281}  \\ \hline
		Standard QAOA & 0.7830  & 0.9402  & 0.7397  & 0.7417  & 0.6880  & 0.4909  & 0.7306  & 0.8733  & 0.8565  & 0.5927  & 0.7742  & 0.8341  & 0.8780  & 0.1508  & 0.7048  & 0.5546  & 0.4302  & 0.5921  \\ 
		RBM QAOA & 0.8258  & 0.9584  & 0.9168  & 0.7344  & 0.3823  & 0.5680  & 0.7310  & 0.6699  & 0.5935  & 0.5029  & 0.5887  & 0.9399  & 0.8693  & 0.7050  & 0.6850  & 0.6323  & 0.6180  & 0.7416  \\ 
		$\mathrm{QAOA}^2$ & 0.7078  & 0.5692  & 0.6646  & 0.6154  & 0.5824  & 0.6154  & 0.6258  & 0.5349  & 0.5678  & 0.5527  & 0.5518  & 0.5185  & 0.6138  & 0.5964  & 0.5569  & 0.5750  & 0.6125  & 0.5789  \\ 
		\textbf{Ours} & 0.8976  & 0.9714  & 0.9339  & 0.8502  & 0.8401  & 0.8374  & \textbf{0.8884}  & 0.9394  & 0.9561  & 0.5527  & \textbf{0.9204}  & 0.9590  & 0.9268  & 0.7621  & 0.8546  & 0.8009  & 0.7842  & \textbf{0.8479}  \\ \hline
		
		\textbf{\multirow{2}*{Time (second))}} & \multicolumn{7}{c|}{$d=2$} & \multicolumn{4}{c|}{$d=3$} & \multicolumn{7}{c|}{$d=4$} \\ \cline{2-19}
		\textbf{} & $i=5$ & $i=6$ & $i=7$ & $i=8$ & $i=9$ & $i=10$ & Avg ($i$) & $i=6$ & $i=8$ & $i=10$ & Avg ($i$) & $i=5$ & $i=6$ & $i=7$ & $i=8$ & $i=9$ & $i=10$ & Avg ($i$) \\ \hline
		Standard QAOA & 47.24  & 52.89  & 80.35  & 97.20  & 294.30  & 892.44  & 244.07  & 63.08  & 135.16  & 1036.18  & 411.47  & 61.66  & 81.27  & 78.93  & 141.96  & 532.55  & 1183.22  & 346.60  \\ 
		RBM QAOA & 31.50  & 41.40  & 86.96  & 80.29  & 160.47  & 333.07  & 122.28  & 102.81  & 250.31  & 456.50  & 269.87  & 51.64  & 74.13  & 133.03  & 259.47  & 297.81  & 507.10  & 220.53  \\ 
		\textbf{Ours} & 13.51  & 18.75  & 31.87  & 50.18  & 98.93  & 310.68  & \textbf{87.32}  & 25.62  & 60.00  & 312.17  & \textbf{132.60}  & 16.79  & 28.56  & 41.28  & 62.10  & 115.53  & 323.13  & \textbf{97.90}  \\ \hline
	\end{tabular}
       \begin{tablenotes}
		\item[] Note: Each experiment was repeated ten times for every graph, and the results were averaged to provide. \textcolor{gray}{Gray} represents the performance of the classical GW method. \textbf{Bold} represents the best AR value of these QAOA methods.
	\end{tablenotes}
\end{table}

\begin{table}[!ht]
	\centering
	\tiny
	\caption{\small{\textbf{Comparison of AR between the baseline methods and the proposed method in weighted complete graphs.} }}\label{tab7}
	\begin{tabular}{|c|cccccc|c|}
		\hline
		\textbf{AR (max)} & $i=5$ & $i=6$ & $i=7$ & $i=8$ & $i=9$ & $i=10$ & Avg ($i$) \\ \hline
		\textcolor{gray}{GW (Classic)} & \textcolor{gray}{1.0000} & \textcolor{gray}{1.0000} & \textcolor{gray}{1.0000}  & \textcolor{gray}{1.0000} & \textcolor{gray}{1.0000} & \textcolor{gray}{1.0000} & \textcolor{gray}{1.0000}  \\ \hline
		Standard QAOA & 0.9678  & 0.9615  & 0.9020  & 0.8713  & 0.9380  & 0.9121  & 0.9255  \\ 
		RBM QAOA & 0.9563  & 0.8803  & 0.9077  & 0.8520  & 0.9573  & 0.9332  & 0.9145  \\ 
		$\mathrm{QAOA}^2$ & 0.5938  & 0.4889  & 0.4861  & 0.4219  & 0.5969  & 0.5750  & 0.5271  \\ 
		Ours & 0.9899  & 0.9710  & 0.9268  & 0.9010  & 0.9760  & 0.9456  & \textbf{0.9517}  \\ \hline
		
		\textbf{AR (avg)} & $i=5$ & $i=6$ & $i=7$ & $i=8$ & $i=9$ & $i=10$ & Avg ($i$) \\ \hline
		\textcolor{gray}{GW (Classic)} & \textcolor{gray}{0.9329}  & \textcolor{gray}{0.9421}  & \textcolor{gray}{0.9437}  & \textcolor{gray}{0.9453}  & \textcolor{gray}{0.9456}  & \textcolor{gray}{0.9474}  & \textcolor{gray}{0.9428}  \\ \hline
		Standard QAOA & 0.9251  & 0.9018  & 0.8369  & 0.7325  & 0.7993  & 0.8296  & 0.8375  \\ 
		RBM QAOA & 0.8139  & 0.7974  & 0.8570  & 0.7676  & 0.8824  & 0.8417  & 0.8266  \\ 
		$\mathrm{QAOA}^2$ & 0.5541  & 0.4468  & 0.3975  & 0.3401  & 0.5838  & 0.5585  & 0.4801  \\
		Ours & 0.9072  & 0.9381  & 0.8968  & 0.8566  & 0.9012  & 0.8990  & \textbf{0.8998}  \\ \hline
		
		\textbf{Time (second)} & $i=5$ & $i=6$ & $i=7$ & $i=8$ & $i=9$ & $i=10$ & Avg ($i$) \\ \hline
		Standard QAOA & 96.61  & 97.60  & 161.26  & 166.80  & 387.26  & 1057.99  & 327.92  \\ 
		RBM QAOA & 115.48  & 143.02  & 229.73  & 324.38  & 366.92  & 453.42  & 272.16  \\ 
		Ours & 15.69  & 26.77  & 43.93  & 77.70  & 135.09  & 401.20  & \textbf{116.73}  \\ \hline
	\end{tabular}
        \begin{tablenotes}
		\item[] Note: Each experiment was repeated ten times for every graph, and the results were averaged to provide. \textcolor{gray}{Gray} represents the performance of the classical GW method. \textbf{Bold} represents the best AR value of these QAOA methods.
	\end{tablenotes}
\end{table}

\begin{table}[!ht]
	\centering
	\tiny
	\caption{\small{\textbf{Comparison between the different \textit{p} layers of PIL QAOA method in unweighted random graphs.} }}\label{tab8}
	\begin{tabular}{|c|cccc|c|cccc|c|cccc|c|}
		\hline
		\textbf{\multirow{2}*{AR (max)}} & \multicolumn{5}{c|}{$i=5$} & \multicolumn{5}{c|}{$i=6$} & \multicolumn{5}{c|}{$i=7$} \\ \cline{2-16}
		~ & $ep=0.2$ & $ep=0.4$ & $ep=0.6$ & $ep=0.8$ & Avg ($ep$)& $ep=0.2$ & $ep=0.4$ & $ep=0.6$ & $ep=0.8$ & Avg ($ep$) & $ep=0.2$ & $ep=0.4$ & $ep=0.6$ & $ep=0.8$ & Avg ($ep$) \\ \hline
		$p=1$ & 0.9975  & 0.9920  & 1.0000  & 0.9767  & 0.9916  & 0.9860  & 0.9683  & 1.0000  & 0.9938  & 0.9870  & 0.9971  & 0.9414  & 0.9456  & 0.9782  & 0.9656  \\ 
		$p=2$ & 1.0000  & 1.0000  & 1.0000  & 0.9783  & 0.9946  & 0.9980  & 1.0000  & 1.0000  & 1.0000  & 0.9995  & 1.0000  & 0.9757  & 0.9867  & 0.9991  & 0.9904  \\ 
		$p=3$ & 1.0000  & 1.0000  & 1.0000  & 1.0000  & 1.0000  & 0.9970  & 0.9994  & 0.9958  & 0.9994  & 0.9979  & 0.9947  & 0.9886  & 0.9931  & 0.9955  & 0.9930  \\ 
		$p=4$ & 0.9975  & 1.0000  & 1.0000  & 1.0000  & 0.9994  & 0.9980  & 0.9917  & 1.0000  & 0.9975  & 0.9968  & 1.0000  & 0.9814  & 1.0000  & 0.9964  & 0.9945  \\ 
		$p=5$ & 1.0000  & 1.0000  & 1.0000  & 0.9967  & 0.9992  & 1.0000  & 1.0000  & 1.0000  & 0.9950  & 0.9988  & 1.0000  & 0.9786  & 1.0000  & 1.0000  & 0.9947  \\ \hline
		
		\textbf{\multirow{2}*{AR (max)}} & \multicolumn{5}{c|}{$i=8$} & \multicolumn{5}{c|}{$i=9$} & \multicolumn{5}{c|}{$i=10$} \\ \cline{2-16}
		~ & $ep=0.2$ & $ep=0.4$ & $ep=0.6$ & $ep=0.8$ & Avg ($ep$)& $ep=0.2$ & $ep=0.4$ & $ep=0.6$ & $ep=0.8$ & Avg ($ep$) & $ep=0.2$ & $ep=0.4$ & $ep=0.6$ & $ep=0.8$ & Avg ($ep$) \\ \hline
		$p=1$ & 0.9550  & 0.9460  & 0.9455  & 0.9550  & 0.9504  & 0.9467  & 0.9446  & 0.9557  & 0.9429  & 0.9475  & 0.9227  & 0.9306  & 0.9372  & 0.9176  & 0.9270  \\ 
		$p=2$ & 0.9773  & 0.9900  & 0.9600  & 0.9950  & 0.9806  & 0.9533  & 0.9569  & 0.9664  & 0.9453  & 0.9555  & 0.9455  & 0.9362  & 0.9450  & 0.9186  & 0.9363  \\ 
		$p=3$ & 0.9951  & 0.9890  & 0.9920  & 0.9912  & 0.9918  & 0.9744  & 0.9601  & 0.9689  & 0.9518  & 0.9638  & 0.9582  & 0.9562  & 0.9450  & 0.9100  & 0.9424  \\ 
		$p=4$ & 0.9988  & 0.9930  & 0.9964  & 0.9793  & 0.9919  & 0.9989  & 0.9869  & 0.9786  & 0.9541  & 0.9796  & 0.9791  & 0.9881  & 0.9509  & 0.9214  & 0.9599  \\
		$p=5$ & 1.0000  & 0.9900  & 1.0000  & 1.0000  & 0.9975  & 0.9978  & 0.9909  & 0.9900  & 0.9747  & 0.9884  & 1.0000  & 0.9969  & 0.9583  & 0.9357  & 0.9727  \\ \hline
		
		\textbf{\multirow{2}*{AR (avg)}} & \multicolumn{5}{c|}{$i=5$} & \multicolumn{5}{c|}{$i=6$} & \multicolumn{5}{c|}{$i=7$} \\ \cline{2-16}
		~ & $ep=0.2$ & $ep=0.4$ & $ep=0.6$ & $ep=0.8$ & Avg ($ep$)& $ep=0.2$ & $ep=0.4$ & $ep=0.6$ & $ep=0.8$ & Avg ($ep$) & $ep=0.2$ & $ep=0.4$ & $ep=0.6$ & $ep=0.8$ & Avg ($ep$) \\ \hline
		$p=1$ & 0.9935 	&0.9872 	&0.9945 	&0.9587 	&0.9835 	&0.9832 	&0.9647 	&0.9984 	&0.9853 	&0.9829 	&0.9889 	&0.9361 &	0.9349 	&0.9720 	&0.9580   \\ 
		$p=2$ & 0.9950 	&0.9920 	&0.9945 	&0.9637 	&0.9863 	&0.9920 	&0.9867 	&0.9964 	&0.9905 	&0.9914 	&0.9957 	&0.9517 &	0.9604 	&0.9727 	&0.9701   \\ 
		$p=3$ & 0.9980 	&0.9976 	&0.9980 	&0.9750 	&0.9922 	&0.9934 	&0.9887 	&0.9895 	&0.9914 	&0.9908 	&0.9931 	&0.9666 &	0.9799 	&0.9576	&0.9743   \\ 
		$p=4$ & 0.9980 	&0.9976 	&0.9980 	&0.9750 	&0.9922 	&0.9934 	&0.9887 	&0.9895 	&0.9914 	&0.9908 	&0.9931 	&0.9666 &	0.9799 	&0.9576	&0.9743   \\ 
		$p=5$ & 0.9990 	&0.9950 	&0.9990 	&0.9873 	&0.9951 	&0.9948 	&0.9913 	&0.9980 	&0.9922 	&0.9941 	&0.9991 	&0.9700 &	0.9931 	&0.9796 	&0.9855   \\ \hline
		
		\textbf{\multirow{2}*{AR (avg)}} & \multicolumn{5}{c|}{$i=8$} & \multicolumn{5}{c|}{$i=9$} & \multicolumn{5}{c|}{$i=10$} \\ \cline{2-16}
		~ & $ep=0.2$ & $ep=0.4$ & $ep=0.6$ & $ep=0.8$ & Avg ($ep$)& $ep=0.2$ & $ep=0.4$ & $ep=0.6$ & $ep=0.8$ & Avg ($ep$) & $ep=0.2$ & $ep=0.4$ & $ep=0.6$ & $ep=0.8$ & Avg ($ep$) \\ \hline
		$p=1$ & 0.9513 	&0.9398 	&0.9384 	&0.9473 	&0.9442 	&0.9302 	&0.9292 	&0.9473 	&0.9119 	&0.9297 	&0.9144 	&0.9157 	&0.9141 	&0.8970 	&0.9103   \\ 
		$p=2$ & 0.9640 	&0.9568 	&0.9431 	&0.9741 	&0.9595 	&0.9393 	&0.9312 	&0.9584 	&0.9206 	&0.9374 	&0.9305 	&0.9261 	&0.9433 	&0.9070 	&0.9267   \\ 
		$p=3$ & 0.9812 	&0.9646 	&0.9776 	&0.9801	&0.9759 	&0.9598 	&0.9466 	&0.9551 	&0.9392	&0.9502 	&0.9409 	&0.9492 	&0.9428 	&0.8711	&0.9260   \\ 
		$p=4$ & 0.9898 	&0.9694 	&0.9743 	&0.9646 	&0.9745 	&0.9800 	&0.9779 	&0.9593 	&0.9400 	&0.9643 	&0.9676 	&0.9770 	&0.9473 	&0.8846 	&0.9442   \\
		$p=5$ & 0.9844 	&0.9788 	&0.9751 	&0.9922 	&0.9826 	&0.9811 	&0.9856 	&0.9567 	&0.9442 	&0.9669 	&0.9698 	&0.9800 	&0.9431 	&0.8835 	&0.9441   \\ \hline
	\end{tabular}
         \begin{tablenotes}
		\item[] Note: Each experiment was repeated ten times for every graph, and the results were averaged to provide.
	\end{tablenotes}
\end{table}

\begin{table}[!ht]
	\centering
	\tiny
	\caption{\small{\textbf{Comparison between the different \textit{p} layers of PIL QAOA method in unweighted regular graphs.} }}\label{tab9}
	\begin{tabular}{|c|cccccc|c|ccc|c|cccccc|c|}
		\hline
		\textbf{\multirow{2}*{AR (max)}} & \multicolumn{7}{c|}{$d=2$} & \multicolumn{4}{c|}{$d=3$} & \multicolumn{7}{c|}{$d=4$} \\ \cline{2-19}
		\textbf{} & $i=5$ & $i=6$ & $i=7$ & $i=8$ & $i=9$ & $i=10$ & Avg ($i$) & $i=6$ & $i=8$ & $i=10$ & Avg ($i$) & $i=5$ & $i=6$ & $i=7$ & $i=8$ & $i=9$ & $i=10$ & Avg ($i$) \\ \hline
		$p=1$ & 1.0000  & 1.0000  & 1.0000  & 0.9950  & 1.0000  & 0.9660  & 0.9935  & 0.9986  & 0.9870  & 0.9438  & 0.9765  & 1.0000  & 1.0000  & 0.9920  & 0.9833  & 0.9371  & 0.9663  & 0.9798  \\ 
		$p=2$ & 1.0000  & 1.0000  & 1.0000  & 1.0000  & 1.0000  & 0.9676  & 0.9946  & 1.0000  & 0.9990  & 0.9438  & 0.9809  & 1.0000  & 1.0000  & 1.0000  & 0.9917  & 0.9471  & 0.9739  & 0.9855  \\ 
		$p=3$ & 1.0000  & 0.9993  & 0.9957  & 1.0000  & 0.9970  & 0.9701  & 0.9937  & 0.9998  & 0.9960  & 0.9512  & 0.9823  & 1.0000  & 1.0000  & 0.9999  & 0.9998  & 0.9481  & 0.9757  & 0.9873  \\ 
		$p=4$ & 1.0000  & 1.0000  & 1.0000  & 1.0000  & 1.0000  & 0.9740  & 0.9957  & 1.0000  & 0.9990  & 0.9560  & 0.9850  & 1.0000  & 1.0000  & 1.0000  & 1.0000  & 0.9567  & 0.9815  & 0.9897  \\ 
		$p=5$ & 1.0000  & 1.0000  & 1.0000  & 1.0000  & 1.0000  & 0.9803  & 0.9967  & 1.0000  & 0.9970  & 0.9637  & 0.9869  & 1.0000  & 1.0000  & 1.0000  & 1.0000  & 0.9700  & 0.9872  & 0.9929  \\ \hline
		
		\textbf{\multirow{2}*{AR (avg)}} & \multicolumn{7}{c|}{$d=2$} & \multicolumn{4}{c|}{$d=3$} & \multicolumn{7}{c|}{$d=4$} \\ \cline{2-19}
		\textbf{} & $i=5$ & $i=6$ & $i=7$ & $i=8$ & $i=9$ & $i=10$ & Avg ($i$) & $i=6$ & $i=8$ & $i=10$ & Avg ($i$) & $i=5$ & $i=6$ & $i=7$ & $i=8$ & $i=9$ & $i=10$ & Avg ($i$) \\ \hline
		$p=1$ & 1.0000  & 1.0000  & 0.9987  & 0.9930  & 0.9965  & 0.9436  & 0.9886  & 0.9957  & 0.9822  & 0.9342  & 0.9707  & 1.0000  & 0.9940  & 0.9880  & 0.9737  & 0.9283  & 0.9597  & 0.9739  \\ 
		$p=2$ & 1.0000  & 1.0000  & 0.9993  & 1.0000  & 0.9988  & 0.9432  & 0.9902  & 0.9977  & 0.9858  & 0.9312  & 0.9716  & 1.0000  & 0.9956  & 0.9976  & 0.9833  & 0.9406  & 0.9570  & 0.9790  \\ 
		$p=3$ & 0.9993  & 0.9985  & 0.9910  & 0.9980  & 0.9860  & 0.9410  & 0.9856  & 0.9991  & 0.9872  & 0.9389  & 0.9751  & 0.9995  & 0.9983  & 0.9986  & 0.9817  & 0.9396  & 0.9591  & 0.9795  \\ 
		$p=4$ & 1.0000  & 1.0000  & 0.9973  & 1.0000  & 0.9980  & 0.9407  & 0.9893  & 1.0000  & 0.9933  & 0.9396  & 0.9776  & 1.0000  & 1.0000  & 1.0000  & 0.9843  & 0.9496  & 0.9648  & 0.9831  \\ 
		$p=5$ & 1.0000  & 1.0000  & 0.9993  & 0.9995  & 0.9985  & 0.9525  & 0.9916  & 1.0000  & 0.9934  & 0.9568  & 0.9834  & 1.0000  & 1.0000  & 1.0000  & 0.9947  & 0.9532  & 0.9663  & 0.9857  \\ \hline
	\end{tabular}
          \begin{tablenotes}
		\item[] Note: Each experiment was repeated ten times for every graph, and the results were averaged to provide.
	\end{tablenotes}
\end{table}

\begin{table}[!ht]
	\centering
	\tiny
	\caption{\small{\textbf{Comparison between the different \textit{p} layers of PIL QAOA method in unweighted complete graphs.} }}\label{tab10}
	\begin{tabular}{|c|cccccc|c|}
		\hline
		\textbf{AR (max)} & $i=5$ & $i=6$ & $i=7$ & $i=8$ & $i=9$ & $i=10$ & Avg ($i$) \\ \hline
		$p=1$ & 1.0000  & 1.0000  & 1.0000  & 0.9944  & 1.0000  & 0.9932  & 0.9979  \\ 
		$p=2$ & 1.0000  & 1.0000  & 1.0000  & 1.0000  & 1.0000  & 0.9992  & 0.9999  \\ 
		$p=3$ & 1.0000  & 0.9998  & 0.9997  & 0.9981  & 0.9997  & 0.9796  & 0.9962  \\
		$p=4$ & 1.0000  & 1.0000  & 1.0000  & 1.0000  & 1.0000  & 0.9980  & 0.9997  \\ 
		$p=5$ & 1.0000  & 1.0000  & 1.0000  & 0.9606  & 0.9840  & 0.8876  & 0.9720  \\ \hline
		
		\textbf{AR (avg)} &$i=5$ & $i=6$ & $i=7$ & $i=8$ & $i=9$ & $i=10$ & Avg ($i$) \\ \hline
		$p=1$ & 1.0000  & 0.9571  & 0.9577  & 0.9133  & 0.9250  & 0.9322  & 0.9475  \\ 
		$p=2$ & 1.0000  & 0.9619  & 0.9728  & 0.9390  & 0.9600  & 0.9564  & 0.9650  \\ 
		$p=3$ & 1.0000  & 0.9996  & 0.9982  & 0.9930  & 0.9859  & 0.9679  & 0.9908  \\ 
		$p=4$ & 1.0000  & 0.9997  & 0.9975  & 0.9525  & 0.9524  & 0.9183  & 0.9701  \\ 
		$p=5$ & 1.0000  & 0.9986  & 0.9950  & 0.9339  & 0.9244  & 0.8814  & 0.9555  \\ \hline
	\end{tabular}
           \begin{tablenotes}
		\item[] Note: Each experiment was repeated ten times for every graph, and the results were averaged to provide.
	\end{tablenotes}
\end{table}

\medskip 
\onecolumngrid

\end{document}